\documentstyle[11pt]{article}
\normalbaselines
\newcounter{eqnletter}[equation]
\catcode`\@=11
\@addtoreset{equation}{section}   % Makes \chapter reset 'equation' counter.
\catcode`\@=12

\begin{document}

\begin{center}

{\LARGE\bf New Fundamental Symmetries of Integrable Systems
and Partial Bethe Ansatz}\footnote{This work was partially
supported by DFG grant No. 436 POL 113/77/0 (S). }

\vskip 1.5cm

{\large {\bf A.G. Ushveridze} }

\vskip 0.4 cm

Department of Theoretical Physics, University of {\L}\'od\'z,\\
Pomorska str. 149/153, 90-236 {\L}\'od\'z, Poland\footnote{E-mail
address: alexush@mvii.uni.lodz.pl and alexush@krysia.uni.lodz.pl} \\
and\\
Institute of Theoretical Physics, Technical University of Clausthal,\\
Arnold Sommerfeld str. 6, 38678 Clausthal-Zellerfeld, Germany\\

\end{center}

\vspace{1 cm}

\begin{abstract}

We introduce a new concept of quasi-Yang-Baxter algebras.
The quantum quasi-Yang-Baxrer algebras 
being simple but non-trivial deformations of
ordinary algebras of monodromy matrices realize a new type 
of quantum dynamical symmetries and find an unexpected and 
remarkable applications in quantum inverse scattering method (QISM).
We show that applying to quasi-Yang-Baxter algebras the
standard procedure of QISM one obtains new wide classes of quantum
models which, being integrable (i.e. having enough number
of commuting integrals of motion) are only quasi-exactly solvable  
(i.e. admit an algebraic Bethe ansatz solution 
for arbitrarily large but limited parts of the spectrum). These
quasi-exactly solvable models naturally arise as
deformations of known exactly solvable ones.
A general theory of such deformations is proposed.
The correspondence ``Yangian --- quasi-Yangian'' and 
``$XXX$ spin models --- quasi-$XXX$ spin models'' is
discussed in detail. We also construct the classical conterparts of
quasi-Yang-Baxter algebras and show that they 
naturally lead to new classes of classical integrable models. 
We conjecture that these models are quasi-exactly solvable
in the sense of classical inverse scattering method, i.e.
admit only partial construction of action-angle variables.

\end{abstract}

\newpage 

\tableofcontents

\newpage

\section{Introduction}
\label{1}

The quantum inverse scattering method (QISM) formulated in
the works \cite{SkFa78,Sk78,Fa79,SkTaFa79,KuSk79,TaFa79} of
the Leningrad group is at present time the
bigest factory of quantum integrable and exactly solvable models most 
of which are of a great physical and mathematical relevance (see e.g. 
refs. \cite{KuSk82,KoBoIz93,Fa95,Fa96} and references therein). 
The main idea of this method rests on 
the use of a special associative algebra ${\cal T}_R$ (known in the 
literature under names of Yang--Baxter algebra, Zamolodchikov algebra,
Fundamental Commutation Relations, etc.)\footnote{Algebra
${\cal T}_R$ appeared first in Yang's papers
\cite{Ya67,Ya68} as algebra of non-relativistic
$S$-matrices and in Baxter's works \cite{Ba72,Ba78} as
algebra of transfer matrices. Later it reappeared in
the papers by Zamolodchikov and Zamolodchikov \cite{ZaZa78,ZaZa79} 
as algebra of relativistic $S$-matrices. The most regular
way of introducing ${\cal T}_R$ algebras was formulated by 
Faddeev's group 
\cite{SkFa78,Sk78,Fa79,SkTaFa79,KuSk79,TaFa79,Sk82,Sk83} 
and also by Drinfeld \cite{Dr85,Dr88} and Jimbo
\cite{Ji85} who proved that ${\cal T}_R$ is a Hopf algebra. }, 
whose generators $T_{\alpha\beta}, \ \alpha,\beta=1,\ldots,d$ 
(forming the so-called {\it monodromy matrix}) satysfy the system of 
bilinear relations
\begin{eqnarray}
R_{\alpha\beta\gamma\delta}(\lambda-\mu)
T_{\gamma\rho}(\lambda)T_{\delta\sigma}(\mu)=
T_{\beta\delta}(\mu)T_{\alpha\gamma}(\lambda)
R_{\gamma\delta\rho\sigma}(\lambda-\mu).
\label{1.1}
\end{eqnarray}
Tensor $R_{\alpha\beta\gamma\delta}(\lambda)$ (which is usually called 
the $R$-matrix) obeys the famous Yang--Baxter equation and completely
determines the structure of algebra ${\cal T}_R$. The role of this 
algebra in QISM can be explained as follows. By using relations 
(\ref{1.1}) one easily constructs the families of operators 
\begin{eqnarray}
H(\lambda)={\cal P}[\{T_{\alpha_i\beta_i}(\lambda_i)\}]
\label{1.2}
\end{eqnarray}
commuting with each other for any values of $\lambda$. Here
${\cal P}[...]$ denote some polynomials in generators 
$T_{\alpha_i\beta_i}(\lambda_i)$ and $\lambda_i$ are some fixed 
functions of $\lambda$. These operators 
are considered as integrals of motion of
some quantum systems. To solve an eigenvalue problem for
these systems one uses a special ansatz
\begin{eqnarray}
\psi={\cal Q}[\{T_{\alpha_i\beta_i}(\xi_i)\}]|0\rangle
\label{1.3}
\end{eqnarray}
(the so-called Bethe ansatz) in which $|0\rangle$ is the so-called 
vacuum vector (playng the role of the lowest weight vector for the 
representation space of algebra ${\cal T}_R$) and ${\cal Q}[...]$ 
are again some polynomials in generators $T_{\alpha_i\beta_i}(\xi_i)$ 
taken at certain points $\xi_i$. The coordinates of these points are 
found from a system of numerical equations known under name of Bethe
ansatz equations. Summarizing, one can say that both the construction 
and solution of quantum problems in QISM is a {\it purely algebraic 
procedure} whose realization requires the knowledge of only two things: 
the defining relations of algebra ${\cal T}_R$ itself and the properties 
of its lowest weight representations. It turns out that most of the 
models obtained and solved in such a way are {\it integrable} 
(i.e. admit enough number of commuting integrals of motion) and 
{\it exactly solvable} (i.e. can be solved algebraically for the whole 
spectrum)\footnote{In quantum mechanics the notions of integrability 
and exact solvability do not coincide because of the absence of
a quantum analogue of the Liouville theorem.}.
This enables one to qualify the ${\cal T}_R$ algebra as a generator of
{\it integrable} and {\it exactly solvable} quantum problems.

\medskip
In this paper  we intend to introduce a new concept 
of ``quasi-${\cal T}_R$ algebras''.
These algebras are distinguished by the fact that ``almost'' the same
construction which led in the case of the ordinary ${\cal T}_R$ 
algebras to exactly solvable models, leads in the case of 
quasi-${\cal T}_R$ algebras to {\it integrable} 
but only {\it quasi-exactly solvable} 
models\footnote{The notion of ``quasi-exact solvability'' was 
introduced in paper \cite{TuUs87}.}, 
i.e. models admitting an algebraic solution only for
some limited parts of the spectrum\footnote{Numerous examples of 
quasi-exactly solvable models and various methods for their 
construction 
can be found in several review 
articles \cite{Us89,Sh89,Us92,UlZa92,Sh94,Tu94,GKO94} 
and in the book \cite{Us94}.}. 
The quasi-${\cal T}_R$ algebra is a very simple but non-trivial 
deformation of the ordinary ${\cal T}_R$ algebra. The form of the 
commutation relations for ${\cal T}_R$ and quasi-${\cal T}_R$ 
algebras ``almost'' coincide and all their properties (including
construction of both commuting integrals of motion and their 
solutions) are ``almost'' the same. However, namely this ``slight'' 
difference makes the quasi-${\cal T}_R$ algebra a new mathematical 
object having many new exciting mathematical properties and a wide 
range of possible physical applications. The simplest
example of quasi-${\cal T}_R$ algebra has already been
considered in our previous paper \cite{Us97}.

\medskip
Quasi-${\cal T}_R$ algebra is not an
absolutely unexpected thing. The feeling that something like this may
exist immediately appears if one remembers
some facts from the 
theory of exactly and quasi-exactly solvable problems.
Indeed, consider for example a pair of two 
simple quantum-mechanical models with hamiltonians
\begin{eqnarray}
H=-\sum_{i=1}^d\frac{\partial^2}{\partial
x_i^2}+\sum_{i=1}^d b_i^2 x_i^2
\label{1.4}
\end{eqnarray}
and
\begin{eqnarray}
H_n=-\sum_{i=1}^d\frac{\partial^2}{\partial
x_i^2}+\sum_{i=1}^d b_i^2 x_i^2 +2ar^2\left[\sum_{i=1}^d b_i x_i^2
-2n-1-\frac{d}{2}\right]+a^2 r^6, \quad n=0,1,\ldots.
\label{1.5}
\end{eqnarray}
The first model is known to everybody. This is an ordinary
$d$-dimensional harmonic oscillator which is {\it integrable}
(since it admits a complete separation of variables) and 
{\it exactly solvable}
(since all its eigenvalues and eigenfunctions can be
constructed algebraically). The second model of a
$d$-dimensional anharmonic oscillator, which appeared on the
physical scene only recently \cite{Us88b} (see also 
\cite{Us89,Us94}), 
is also {\it integrable} (because its variables can be separated 
in the
generalized ellipsoidal coordinates \cite{Us88c}) but, in
contrast with model (\ref{1.4}), 
it is only {\it quasi-exactly solvable} (since
admits only partial algebraic solution of spectral problem
for any non-negative integer $n$)\footnote{Indeed, it can be 
explicitly shown that, for any given $n$,
the number of algebraically constructable solutions of
model (\ref{1.5}) is only $(n+d)!/(n!d!)$.}. 

\medskip
What can we learn from comparing these two models?
First of all, we see that model (\ref{1.5}) can be
considered as a {\it deformation} of the model (\ref{1.4}) and 
the role of
the deformation parameter is played by $a$. 
Second, it is clear that this deformation 
preserves the integrability property.
Third, this deformation is of the splitting type,
because it transforms a single 
exactly solvable model into an infinite sequence of
quasi-exactly solvable models with different hamiltonians
and different number of exact solutions \cite{Us94}.

\medskip
It turns out that this situation is quite general. In fact,
it can be shown that any integrable and exactly solvable model 
satisfying some simple and rather general conditions can be
constructively deformed into an infinite sequence of
integrable and quasi-exactly solvable models \cite{Us94}. 
In particular, this is true for models associated with various 
${\cal T}_R$ algebras and obtainable by means
of QISM\footnote{The case of the simplest
models of such a sort (the so-called Gaudin
models \cite{Ga83}) was considered in refs. \cite{Us92,Us94}. 
For more
general discussion of this subject see
sections \ref{15} and \ref{16} of present paper.}.
But if so, may be this deformation can be understood on a
purely {\it algebraic} level? May be, there exists some
non-trivial deformation of the underlying ${\cal T}_R$ algebra, 
which
could be called ``quasi-${\cal T}_R$ algebra'', and which would
naturally lead to quasi-exactly solvable versions of
integrable models?

\medskip
The aim of this paper is to give a positive answer to this question.
To make the text more readable, we devote the main attention to the 
simplest
rational ${\cal T}_R$-algebra corresponding to a monodromy matrix of
the size $2\times 2$. The rationality means that both
generators $T_{\alpha\beta}(\lambda), \ \alpha,\beta=1,2$ and tensor 
$R_{\alpha\beta\gamma\delta}(\lambda),\ \alpha,\beta,\gamma,\delta=1,2$ 
are rational functions of $\lambda$. Such an algebra is called the
Yangian ${\cal Y}[sl(2)]$ and is characterized by the
following choice of non-zero elements of tensor 
$R_{\alpha\beta\gamma\delta}(\lambda)$
\begin{eqnarray}
R_{1111}(\lambda)=R_{2222}(\lambda)&=&1,\nonumber\\
R_{1212}(\lambda)=R_{2121}(\lambda)&=&\lambda/(\lambda+\eta),
\nonumber\\
R_{1221}(\lambda)=R_{2112}(\lambda)&=&\eta/(\lambda+\eta),
\label{1.6}
\end{eqnarray}
where $\eta$ is a parameter. For $\eta\neq 0$ the Yangian ${\cal
Y}[sl(2)]$ is a non-commutative and non-cocommutative Hopf 
algebra \cite{Dr85,Dr88}.
In the limit $\eta\rightarrow 0$ it degenerates
into a special (infinite-dimensional) 
Lie -- Gaudin algebra \cite{Ga83} which
we denote by ${\cal G}[sl(2)]$. 
The completely integrable and exactly solvable models
associated with algebras ${\cal G}[sl(2)]$ and ${\cal Y}[sl(2)]$ are
respectively known under names of Gaudin and $XXX$ spin models.

\medskip
The paper is organized as follows. After short exposition of
properties of ordinary algebras ${\cal G}[sl(2)]$ and ${\cal
Y}[sl(2)]$ (sections \ref{2} and \ref{8}) we explain the reader what
we mean under quasi-${\cal G}[sl(2)]$ and quasi-${\cal Y}[sl(2)]$ 
algebras and present their defining relations 
(sections \ref{3} and \ref{9}).
We also give a proof of the existence of these
algebras by contructing their explicit realizations (sections \ref{5} 
and \ref{11}).
In particular, we demonstrate that they actually can be considered as
deformations of algebras ${\cal G}[sl(2)]$ and ${\cal Y}[sl(2)]$
and show that these deformations are of the splitting type, i.e. the
generators of ``quasi'' algebras contain
an additional subscript $n$. The non-triviality lies
however in the way in which this subscript appears in the
defining relations (\ref{1.1}) which, as for the rest, are the same
as in the undeformed case. We show also that the proposed
deformation preserves the 
integrability property (sections \ref{3} and \ref{9}).
Following general prescriptions of QISM,
we construct the infinite series of the associated integrable and
quasi-exactly solvable 
models (sections \ref{3},\ref{6} and \ref{9},\ref{12})
and present their partial solutions in the framework of the
algebraic Bethe ansatz (sections 4 and 10).
We call the models obtained in such a way the quasi-Gaudin
and quasi-$XXX$ models, respectively.

\medskip
In sections 
\ref{13} -- \ref{16} we discuss the obtained results from the
point of view of the existing methods in the theory of
quasi-exactly solvable systems. We consider three general
methods known under names of ``partial algebraization method'' 
(section \ref{14}), ``inverse method of separation of 
variables'' (section \ref{15}) and ``projection method'' 
(section \ref{16}), and show that, in
contrast with the first and second methods, which 
are either not applicable or only partially applicable 
to the study of quasi-${\cal T}_R$ algebras, 
the third one enables one to reproduce correctly
practically all features of the proposed formalizm. However, the
mathematical technique used in this method is so
cumbersome and model-dependent that it hardly can be
practically used beyond the simplest cases of algebras
${\cal G}[sl(2)]$ and ${\cal Y}[sl(2)]$ and their ``quasi''
analogues. 

\medskip
Fortunately, the ``quasi-QISM'' formalism which we
propose in this paper is free of all these deficiences.
Indeed, in sections \ref{17} -- \ref{21} we show that this
formalism is quite 
general and applicable not only to ${\cal G}[sl(2)]$
and ${\cal Y}[sl(2)]$ 
algebras, but also to all ${\cal T}_R$ algebras,  
irrespective of their dimension and a concrete form of
tensor $R_{\alpha\beta\gamma\delta}(\lambda)$. For this we will
use a model independent formalizm based on the notion of
$Z^r$-graded unital associative algebras (which in this
paper we call simply $A$-algebras)
and construct quasi-deformations immediately for the latter. 
The obtained 
results are automatically applicable to any ${\cal T}_R$
algebras since they can be considered as particular cases
of $A$-algebras. This gives us the hope that using this general
deformation technique we will find a lot of new quasi-exactly 
solvable
models interesting both from physical and mathematical
points of view. 

\medskip
One of questions discussed in this
paper (see sections \ref{7} and \ref{20}) is related to 
a special limit of quasi-${\cal T}_R$ algebras when 
the so-called ``spectral parameter'' $\lambda$ 
characterizing their generators tends to infinity. 
We know that in the
case of ordinary (classical or quantum) 
${\cal T}_R$ algebras such a limit leads
respectively to ordinary finite-dimensional Lie algebras or
their $q$-deformed (Lie$_q$) versions. It turns out that 
performing
the same limiting 
procedure with quasi-${\cal T}_R$ algebras we also
obtain some new interesting mathematical structures which can
be called quasi-Lie and quasi-Lie$_q$ algebras.

\medskip
In section \ref{22} we introduce the concept of classical
quasi-${\cal T}_R$ algebras. These algebras naturally arise as the
result of {\it dequantization} of quantum quasi-${\cal T}_R$
algebras discussed in preceding sections. Their generators
are realized as functions on the phase space and obey the
special classical commutation relations. The latter are
expressed through a special non-standard (quasi-Poisson)
bracket which can be considered as a deformation of 
the ordinary Poisson bracket. We show that the the role of classical 
quasi-${\cal T}_R$ algebras in the classical inverse
scattering method (CISM) is the same as that of quantum
quasi-${\cal T}_R$ algebras in QISM. The classical models
associated with classical quasi-${\cal T}_R$ algebras are
integrable but it seems that they are only partially
solvable in the framework of CISM.

\medskip
In the concluding section \ref{23} we give a list of
problems which naturally arise from the results of the
present paper and, in our opinion, deserve a very careful study.

\section{Gaudin algebra and Gaudin model}
\label{2}

In this section we remind the reader some basic facts concerning 
Gaudin algebras, 
their representations and properties of the associated
Gaudin models. More detailed exposition of this subject can
be found in refs. \cite{Ga83,Ju89,Us92,Us94,FeFrRe94}.

\medskip
As we mentioned in Introduction, the Gaudin
algebra ${\cal G}[sl(2)]$ can be obtained from the Yangian 
${\cal Y}[sl(2)]$ (which is characterized by $R$-matrix
of the form (\ref{1.6}) and commutation relations (\ref{1.1}) ) 
in the limit when $\eta\rightarrow 0$.
For this the 
generators $T_{\alpha\beta}(\lambda),\ \alpha,\beta=1,2$
should be taken in the form
\begin{eqnarray}
T_{11}(\lambda)&=&1+\eta S^0(\lambda)+{\cal O}(\eta^2),\nonumber\\
T_{22}(\lambda)&=&1-\eta S^0(\lambda)+{\cal O}(\eta^2),\nonumber\\
T_{12}(\lambda)&=&-\eta S^-(\lambda)+{\cal O}(\eta^2),\nonumber\\
T_{11}(\lambda)&=&+\eta S^+(\lambda)+{\cal O}(\eta^2),
\label{2.1}
\end{eqnarray}
where $S^0(\lambda)$, $S^-(\lambda)$ and $S^+(\lambda)$
are just the generators of ${\cal G}[sl(2)]$. These generators,
which are respectively called neutral, lowering and raising
generators, obey the following commutation relations
\begin{eqnarray}
S^0(\lambda)S^0(\mu)-S^0(\mu)S^0(\lambda)&=&0, \nonumber\\
S^+(\lambda)S^+(\mu)-S^+(\mu)S^+(\lambda)&=&0, \nonumber\\
S^-(\lambda)S^-(\mu)-S^-(\mu)S^-(\lambda)&=&0, \nonumber\\
S^0(\lambda)S^+(\mu)-S^+(\mu)S^0(\lambda)&=&
-(\lambda-\mu)^{-1}\{S^+(\lambda)-S^+(\mu)\},\nonumber\\
S^0(\lambda)S^-(\mu)-S^-(\mu)S^0(\lambda)&=&
+(\lambda-\mu)^{-1}\{S^-(\lambda)-S^-(\mu)\},\nonumber\\
S^-(\lambda)S^+(\mu) -S^+(\mu)S^-(\lambda)&=&
-2(\lambda-\mu)^{-1}\{S^0(\lambda)-S^0(\mu)\}.
\label{2.2}
\end{eqnarray}
From (\ref{2.2}) it immediately follows that operators
\begin{eqnarray}
H(\lambda)=S^0(\lambda)S^0(\lambda)-\frac{1}{2}
S^-(\lambda)S^+(\lambda)-\frac{1}{2}S^+(\lambda)S^-(\lambda)
\label{2.3}
\end{eqnarray}
form a commutative family,
\begin{eqnarray}
[H(\lambda),H(\mu)]=0,
\label{2.4}
\end{eqnarray}
and thus can 
be interpreted as integrals of motion of a certain quantum 
completely integrable model\footnote{Using concrete
realizations of 
generators of algebra ${\cal G}[sl(2)]$, one can easily 
check that the number of independent operators belonging 
to the class $H(\lambda)$ is sufficiently large 
for claiming that this model is
completely integrable.}.
The latter is known under name of Gaudin model.

\medskip
The role of the carrier space in which the
operators $H(\lambda)$ 
act\footnote{We try to avoid the use of the term 
``Hilbert space'' because 
we do not intend to discuss in this paper the 
hermiticity properties of operators $H(\lambda)$ and the 
normalizability of their eigenstates.}  is played by the 
representation space of Gaudin algebra.
In order to construct it one needs to fix the lowest weight vector
$|0\rangle$ and the lowest weight function $F(\lambda)$
obeying the relations
\begin{eqnarray}
S^0(\lambda)|0\rangle=
F(\lambda)|0\rangle, \qquad S^-(\lambda)|0\rangle=0.
\label{2.5}
\end{eqnarray}
After this, we can define the representation space as a 
linear hull of vectors
\begin{eqnarray}
|\xi_1,\ldots,\xi_n\rangle=S^+(\xi_n)\ldots S^+(\xi_1)|0\rangle,
\label{2.6}
\end{eqnarray}
with arbitrary $n$ and $\xi_1,\ldots,\xi_n$. We shall
denote this space by $W_{F(\lambda)}$.

\medskip
The ``Schr\"odinger equation'' for the Gaudin model reads now
\begin{eqnarray}
H(\lambda)\phi=E(\lambda)\phi, \quad \phi\in W_{F(\lambda)}.
\label{2.7}
\end{eqnarray}
The beauty of this equation lies in the fact that all
its solutions can be obtained algebraically within the so-called 
Bethe ansatz
\begin{eqnarray}
\phi=S^+(\xi_n)S^+(\xi_{n-1})\ldots S^+(\xi_2)S^+(\xi_1)|0\rangle,
\label{2.8}
\end{eqnarray}
in which $n$ is an arbitrarily fixed non-negative integer and
$\xi_1,\ldots,\xi_n$ are some still unknown numbers. 
To demonstrate this, one needs to act on Bethe vector by
operator $H(\lambda)$ and, using commutation relations
\begin{eqnarray}
H(\lambda)S^+(\xi_i)
=S^+(\xi_i)H(\lambda)+
2\frac{S^+(\xi_i)S^0(\lambda)
-S^+(\lambda)S^0(\xi_i)}{\lambda-\xi_i}
\label{2.9}
\end{eqnarray}
which elementary 
follow from (\ref{2.2}), transfer $H(\lambda)$ to the
right. The neutral operators $S^0(\lambda)$ and $S^0(\xi_i)$ 
appearing in the right hand
side of (\ref{2.9}) also should be transfered to the right.
For this the fourth relation in the list (\ref{2.2}) can be used.
Finally, we obtain a number of terms with operators 
$H(\lambda)$, $S^0(\lambda)$ and $S^0(\xi_i),\
i=1,\ldots, n$ acting immediately on the lowest weight vector 
$|0\rangle$. After this one
should get rid of these operators by using formulas (\ref{2.5})
together with the relation
\begin{eqnarray}
H(\lambda)|0\rangle=(F^2(\lambda)+F'(\lambda))|0\rangle
\label{2.10}
\end{eqnarray}
which elementary 
follows from (\ref{2.5}) and (\ref{2.3})\footnote{In
order to derive (\ref{2.10}) one should first take
the last relation in the list (\ref{2.2}) in the limit when 
$\mu\rightarrow \lambda$, and using it rewrite the
operator (\ref{2.3}) in the form 
$H(\lambda) = [S^0(\lambda)]^2 + 
[S^0(\lambda)]' - S^+(\lambda)S^-(\lambda)$.
After this one should act by this operator on the 
vacuum vector and then use formulas (\ref{2.5}).}.
The result will consist of
two parts. The first part will be proportional to the Bethe
vector $\phi_n$, while the second part will contain $n$ terms
not having the initial Bethe form (these are the so-called
``unwanted terms''). The condition of cancellation of these
terms imposes some special conditions on numbers 
$\xi_i,\ i=1,\ldots,n$
which up to now were considered as free parameters. 
An explicit form of these conditions, which are commonly
known as Bethe ansatz equations, reads
\begin{eqnarray}
\sum_{k=1,k\neq i}^n \frac{1}{\xi_i-\xi_k}+F(\xi_i)=0,\quad
i=1,\ldots, n.
\label{2.11}
\end{eqnarray}
Putting all these facts together, one can easily obtain 
the final solution 
of the Gaudin spectral problem:
\begin{eqnarray}
E(\lambda)=F^2(\lambda)+ F'(\lambda)+2\sum_{i=1}^n 
\frac{F(\lambda)-F(\xi_i)}{\lambda-\xi_i}.
\label{2.12}
\end{eqnarray}
It is obvious that if $F(\lambda)$ is a rational
function, then the number of solutions of Bethe ansatz equations 
(\ref{2.11}) is finite for any $n$. But since $n$ is an
arbitrary non-negative integer, the complete number of
solutions of problem (\ref{2.7}) is infinite. It can be
shown that these solutions expire all possible solutions of
problem (\ref{2.7}) and therefore it can be 
qualified as exactly solvable.

\section{Quasi-Gaudin algebra and quasi-Gaudin models}
\label{3}

Let us now try to explain the reader what we mean under
words ``quasi-Gaudin algebra''. Note that this will be
namely an attempt of an explanation and not a list of
rigorous definitions and theorems. A strict mathematical
interpretation of formulas of this section will be given 
later in sections \ref{17} -- \ref{21}. 

\medskip
Let $S_n^0(\lambda)$, $S_n^-(\lambda)$ and $S_n^+(\lambda)$
denote the operators parametrized by an integer
number $n$ and obeying the relations
\begin{eqnarray}
S_n^0(\lambda)S_n^0(\mu)-S_n^0(\mu)S_n^0(\lambda)&=&0, \nonumber\\
S_{n+1}^+(\lambda)S_n^+(\mu)-S_{n+1}^+(\mu)S_n^+(\lambda)&=&0, 
\nonumber\\
S_{n-1}^-(\lambda)S_n^-(\mu)-S_{n-1}^-(\mu)S_n^-(\lambda)&=&0, 
\nonumber\\
S_{n+1}^0(\lambda)S_n^+(\mu)-S_n^+(\mu)S_n^0(\lambda)&=&
-(\lambda-\mu)^{-1}\{S_n^+(\lambda)-S_n^+(\mu)\},\nonumber\\
S_{n-1}^0(\lambda)S_n^-(\mu)-S_n^-(\mu)S_n^0(\lambda)&=&
+(\lambda-\mu)^{-1}\{S_n^-(\lambda)-S_n^-(\mu)\},\nonumber\\
S_{n+1}^-(\lambda)S_n^+(\mu) -S_{n-1}^+(\mu)S_n^-(\lambda)&=&
-2(\lambda-\mu)^{-1}\{S_n^0(\lambda)-S_n^0(\mu)\}.
\label{3.1}
\end{eqnarray}
We consider (\ref{3.1}) as the definig relations for
a certain infinite-dimensional algebra which is very similar to
the Gaudin algebra but, in contrast with the latter, it {\it is
not} a Lie algebra. We 
shall call it ``quasi-${\cal G}[sl(2)]$ algebra''
and the expressions staying in the left hand sides of
relations (\ref{3.1}) we refer to as ``quasi-commutators''.
A most unusual feature of algebra ${\cal G}[sl(2)]$
is that it is {\it partially free} i.e. does not contain
quasi-commutation relations between generators whose
indices differ more than by one. One can also say that this
alhebra is {\it local} in the space of
a discrete parameter $n$. 
Note that the form of quasi-commutation relations is invariant
under special ``quasi-similarity'' transformations 
\begin{eqnarray}
S_n^-(\lambda)&\rightarrow&U_{n-1}S_n^-(\lambda)U_n^{-1},\nonumber\\
S_n^0(\lambda)&\rightarrow&U_nS_n^0(\lambda)U_n^{-1},\nonumber\\
S_n^+(\lambda)&\rightarrow&U_{n+1}S_n^+(\lambda)U_n^{-1},
\label{3.1a}
\end{eqnarray}
in which $U_n$ are arbitrary invertible operators.

\medskip
One of the remarkable 
properties of quasi-${\cal G}[sl(2)]$ algebra is that
its generators 
(exactly as in the ${\cal G}[sl(2)]$ algebraic case) can be 
used for building hamiltonians of some completely integrable
quantum systems. 
Indeed, using quasi-commutation relations (\ref{3.1}) it is
not difficult to prove that the operators 
\begin{eqnarray}
H_n(\lambda)=S_n^0(\lambda)S_n^0(\lambda)-\frac{1}{2}
S_{n+1}^-(\lambda)S_n^+(\lambda)-
\frac{1}{2}S_{n-1}^+(\lambda)S_n^-(\lambda)
\label{3.2}
\end{eqnarray}
form commutative families for any $n$:
\begin{eqnarray}
[H_n(\lambda),H_n(\mu)]=0.
\label{3.3}
\end{eqnarray}
At the same time, the operators $H_n(\lambda)$ and $H_m(\mu)$ 
with different $n$ and $m$ do not generally commute with
each other because of 
the absence of appropriate (non-local) relations
in algebra ${\cal G}[sl(2)]$. This means that formula (\ref{3.2}) 
describes in fact an infinite set of {\it different} integrable 
models\footnote{Later, in 
section \ref{5}, after constructing a concrete 
realization of algebra 
${\cal G}[sl(2)]$ we will be able to demonstrate that operators
$H_n(\lambda)$ and 
$H_m(\mu)$ actually do not commute with each other. 
At the same time, the 
number of independent operators contained in the family 
$H_n(\lambda)$ with a given $n$ is sufficiently large to claim that
the corresponding models are completely integrable.}. 
This is, may be, the most important difference with the Gaudin case
where an analogous construction leads to a single completely
integrable model --- the Gaudin model. 

\medskip
Exactly as in the case of Gaudin algebra, the final
construction of models with hamiltonians (\ref{3.2}) 
needs the specification of a carrier space in which
these hamiltonians act. It would be natural to identify it
with the ``representation space'' of 
our algebra\footnote{Generally speaking, the object
which we intend to call the ``representation space'' is not 
a true representation space for 
quasi-${\cal G}[sl(2)]$ algebra since, 
even after 
choosing an appropriate basis in it, we cannot represent any 
element of this 
algebra in a matrix form. This is possible only for some 
special realizations of 
quasi-${\cal G}[sl(2)]$, one of which will be considered in
section \ref{5}. }. The latter can 
be constructed almost in the same way as in the
Gaudin case. 
There is however an essential difference which lies in
the fact that the quasi-analogoues of relations (\ref{2.5}) 
cannot be written 
simultaneously for all ``neutral'' and ``lowering''
operators $S_n^0(\lambda)$ 
and $S_n^-(\lambda)$, i.e. for any $n$. This
simply would lead to an overdetermined system of equations 
for the vacuum vector $|0\rangle$. Roughly speaking, to
have the same number of equations for $|0\rangle$ as in
(\ref{2.5}), one should restrict ourselves to one
arbitrarily fixed value of $n$. 
Taking for example $n=0$, we obtain
\begin{eqnarray}
S_0^0(\lambda)|0\rangle=F(\lambda)|0\rangle,\qquad
S_0^-(\lambda)|0\rangle=0,
\label{3.4}
\end{eqnarray}
where $F(\lambda)$ is, as before, a certain function
playing the role of the lowest weight. 
The imposibility of writing the relations
like (\ref{3.4}) for any $n$ is another unusual property
of quasi-${\cal G}[sl(2)]$ algebra, and it, strictly
speaking, should be postulated from very 
beginning\footnote{Later, 
in section \ref{5} we demonstrate that the lowest
weight vector $|0\rangle$ is actually not annihilated by the 
``lowering operators'' $S_n^-(\lambda)$ if $n\neq 0$. 
Here, however, we 
suggest the reader to consider this fact as a postulate.}.
The presence of such a postulate will play a very important role 
in our further considerations.

\medskip
After fixing the ``lowest weight vector'' $|0\rangle$
together with the 
corresponding ``lowest weight function'' $F(\lambda)$,
we can define the ``representation space'' by analogy
with (\ref{2.6}) i.e. as a linear hull of vectors obtained 
after action of operators $S_n^+(\xi_i)$ on $|0\rangle$. 
It is however intuitivelly clear that there is a big freedom in 
choosing such vectors because the indices of raising operators may
take infinitely many different values. In order to restrict this 
freedom (which, unfortunately, we cannot control at present time), 
we need a ``guiding principle'', which, in particular, 
could be based on a selectional use of some symmetries.
Looking for example at the vectors (\ref{2.6}) forming the 
representation space of ordinary Gaudin algebra, we see that,
due to the commutativity of raising generators
$S^+(\xi_1),\ldots,S^+(\xi_n)$, they are symmetric 
under all permutations of numbers $\xi_1,\ldots,\xi_n$. 
It is natural to demand the same symmetry in the case of 
quasi-Gaudin algebras. Now, however, the raising operators do 
not commute anymore, they only ``quasi-commute'',
and therefore the only way of having such a symmetry
is to choose the corresponding vectors in the form
\begin{eqnarray}
|\xi_1,\ldots,\xi_m\rangle_k=S_{m+k}^+(\xi_m)S_{m-1+k}^+(\xi_{m-1})
\ldots S_{2+k}^+(\xi_2) S_{1+k}^+(\xi_1)|0\rangle
\label{3.5}
\end{eqnarray}
with arbitrary $k\in Z$, $m\in N$ and $\xi_1,\ldots,\xi_m$.
As before, we denote the ``representation 
space'' of algebra (\ref{3.1})
defined in such a way by $W_{F(\lambda)}$. 

\medskip
Now we can complete the construction of integrable models (\ref{3.2})
associated with algebra (\ref{3.1}). We postulate that 
``Schr\"odinger equations'' for these models read
\begin{eqnarray}
H_n(\lambda)\phi=E(\lambda)\phi, \quad \phi\in
W_{F(\lambda)}, \qquad n=0,1,2,\ldots.
\label{3.6}
\end{eqnarray}
Hereafter we shall call such models the ``quasi-Gaudin models''.

\section{Bethe ansatz for quasi-Gaudin models}
\label{4}

It turns out that, in full analogy with the Gaudin case, the
solutions of spectral equations (\ref{3.6}) can 
be found by means of the
algebraic Bethe ansatz. But before constructing the
appropriate Bethe vectors one should discuss some auxiliary
problems concerning the properties of quasi-Gaudin algebra.

\medskip
It is known that in the Gaudin case the lowest weight
vector is always a solution of the Gaudin spectral problem
(see for example formula (\ref{2.10})).
Is this true in the case of quasi-Gaudin models?
In order to answer this question, one should simply act by the
operator $H_n(\lambda)$ on the vector $|0\rangle$ and look
at the result. For this it is convenient to rewrite the
operator $H_n(\lambda)$ in a little bit different form.
Using the last formula in the list (\ref{3.1}) and taking in it
the limit $\mu\rightarrow\lambda$ we obtain
\begin{eqnarray}
S_{n+1}^-(\lambda)S_n^+(\lambda)=S_{n-1}^+(\lambda)S_n^-(\lambda) 
-2\frac{\partial}{\partial \lambda} S_n^0(\lambda).
\label{4.1}
\end{eqnarray}
Substituting (\ref{4.1}) into expression (\ref{3.2}) for 
$H_n(\lambda)$, we get the needed expression for $H_n(\lambda)$
\begin{eqnarray}
H_n(\lambda)=S_n^0(\lambda)S_n^0(\lambda)+
\frac{\partial}{\partial\lambda} S_n^0(\lambda)
-S_{n-1}^+(\lambda)S_n^-(\lambda).
\label{4.2}
\end{eqnarray}
Remembering now that for $n\neq 0$ the vacuum vector
$|0\rangle$ {\it is not} in general 
annihilated by the lowering operators $S_n^-(\lambda)$
and {\it is not} an eigenvector of neutral operators
$S_n^0(\lambda)$,  we can conclude that it
may be an eigenvector of (\ref{4.2}) only if $n=0$. 
Using formulas (\ref{3.4}), we obtain for this case the relation
\begin{eqnarray}
H_0(\lambda)|0\rangle=
\left(F^2(\lambda)+F'(\lambda)\right)|0\rangle
\label{4.3}
\end{eqnarray}
which is similar to the Gaudin relation (\ref{2.10}).

\medskip
Let us now return to the main problem of solving equations (\ref{3.6})
by means of Bethe ansatz. It is natural to take the 
Bethe vector in the form
\begin{eqnarray}
\phi=S_{m+k}^+(\xi_m)S_{m-1+k}^+(\xi_{m-1})
\ldots S_{2+k}^+(\xi_2) S_{k+1}^+(\xi_1)|0\rangle
\label{4.4}
\end{eqnarray}
with some $k\in Z$ and $m\in N$, and, using the relations
\begin{eqnarray}
H_n(\lambda)S_{n-1}^+(\xi_n)=
S_{n-1}^+(\xi_n)H_{n-1}(\lambda) +
2\frac{S_{n-1}^+(\xi_n)S_{n-1}^0(\lambda)
-S_{n-1}^+(\lambda)S_{n-1}^0(\xi_n)}{\lambda-\xi_n},
\label{4.5}
\end{eqnarray}
which follow from the basic relations (\ref{3.1}),
try to transfer the operator $H_n(\lambda)$ to the right
(exactly as in the ordinary Gaudin case).
From formula (\ref{4.5}) it is seen that, in order to start
performing this procedure, it is necessary to take 
\begin{eqnarray}
m+k=n-1. 
\label{4.6}
\end{eqnarray}
Otherwise we would be unable to transfer the operator $H_n(\lambda)$
to the right because of the absence of the corresponding
commutation relations in (\ref{4.5}).
From formula (\ref{4.5}) it is readily seen that permutation 
of the operator $H_n(\lambda)$ with the first raising
generator $S_{n-1}^+(\xi_n)$ of Bethe vector 
replaces $H_n(\lambda)$ by $H_{n-1}(\lambda)$.
After this, $H_{n-1}(\lambda)$ will appear in front of the
next raising generator $S_{n-2}^+(\xi_{n-1})$, and we can
again permute these operators by means of the same
formula (\ref{4.5}) but with $n$ replaced by $n-1$. 
Obviously, this procedure can be
continued further. The same relates to the neutral generators 
$S_{n-1}^0(\lambda)$ and $S_{n-1}^0(\xi_n)$ appearing in
the right hand side of (\ref{4.5}) and thus acting
immediately on $S_{n-2}^+(\xi_{n-1})$. The fourth relation
in the list (\ref{3.1}) enables us to perform the
permutation of these operators, after which the neutral 
generators appear in front of the generator $S_{n-3}^+(\xi_{n-2})$
and will have the index $n-2$. This enables one to use again 
the forth formula in (\ref{3.1}), etc.
Summarising, one can say that any permutation of $H$- and 
$S^0$-operators with
raising generators forming the Bethe vector decreases their
indices by one. Finally, when all the transferences will be
completed, we obtain a number of terms with operators 
$H_{n-m}(\lambda)$, $S_{n-m}^0(\lambda)$ and 
$S_{n-m}^0(\xi_i),\ i=1,\ldots,n$ 
acting immediately on the lowest weight vector $|0\rangle$. 
The standard presciptions to Bethe ansatz technique (see
e.g. section \ref{2}) imply that 
all these operators should be absorbed by 
this vector. But such an absorbtion is
possible only if the lowest weight vector is an
eigenvector of these operators, which, 
according to the previous results, is possible 
only if
\begin{eqnarray}
n-m=0. 
\label{4.7}
\end{eqnarray} 
Comparing formulas (\ref{4.6}) and (\ref{4.7}) we can conclude
that the only case when the
ansatz (\ref{4.4}) for equation (\ref{3.6}) may lead to
some algebraic solutions corresponds to the choice
\begin{eqnarray}
k=-1,\quad m=n.  
\label{4.8}
\end{eqnarray} 
After using the restrictions (\ref{4.8}) the ansatz (\ref{4.4}) 
takes the following final form
\begin{eqnarray}
\phi=S_{n-1}^+(\xi_n)S_{n-2}^+(\xi_{n-1})
\ldots S_1^+(\xi_2) S_0^+(\xi_1)|0\rangle.
\label{4.9}
\end{eqnarray}
Now it remains only to check that it actually contains 
solutions of the quasi-Gaudin spectral problem
(\ref{3.6}). This can be demonstrated exactly 
in the same way as in the
ordinary Gaudin case. Getting rid of the operators
$H_0(\lambda)$, $S_0^0(\lambda)$ and $S_0^0(\xi_i)$,
$i=1,\ldots,n$ standing in front of the vacuum vector by
means of formulas (\ref{3.4}) and (\ref{4.3}),
we obtain two sorts of terms. One of these terms 
will be proportional to the Bethe
vector $\phi$, while the other ones will represent the good
old ``unwanted terms''. It is not difficult to show that the
condition of cancellation of the latter reads
\begin{eqnarray}
\sum_{k=1,k\neq i}^n \frac{1}{\xi_i-\xi_k}+F(\xi_i)=0,\quad
i=1,\ldots, n,
\label{4.10}
\end{eqnarray}
and the final expression for the eigenvalues $E(\lambda)$ is
given by
\begin{eqnarray}
E(\lambda)=F^2(\lambda)+F'(\lambda)+2\sum_{i=1}^n
\frac{F(\lambda)-F(\xi_i)}{\lambda-\xi_i}.
\label{4.11}
\end{eqnarray}
This gives us the desired algebraic solution of problem (\ref{3.6}).
From the above consideration it is clearly seen that 
models with hamiltonians $H_n(\lambda)$ are typical 
quasi-exactly solvable ones.
Indeed, the linear hull 
of all admissible Bethe vectors (\ref{4.9}) forms
only a certain small subspace in the space of vectors
of the general form (\ref{4.4}). The number of explicit
solutions of problems (\ref{3.6}) is finite for any given
$n$ if $F(\lambda)$ is a rational function. At the same
time the dimension of the space $W_{F(\lambda)}$ is
generally infinite-dimensional.
It is interesting that we obtained exactly 
the same set of solutions as in the
case of ordinary Gaudin model. But now these solutions are
distributed between different models.

\section{The existence of quasi-Gaudin algebras} 
\label{5}

Everything what we said in sections \ref{3} and \ref{4} 
can be checked by means of direct calculations.
The only thing which may seem dubious is a very existence
of an algebra with such strange properties.
In order to make sure that it does actually exist, one
should construct at least one of its explicit realizations.
One of possible realization can be constructed 
immediately from the
generators of Gaudin algebra. Let us take the generators
$S^0(\lambda), S^-(\lambda), S^+(\lambda)$
introduced in section \ref{2} and assume that they act in the
representation space characterized by the lowest weight
function $F^0(\lambda)$. 
Let us also introduce a special limiting operator
\begin{eqnarray}
S^0=\lim_{\lambda\rightarrow\infty}\lambda S^0(\lambda).
\label{5.1}
\end{eqnarray}
Then the desired realization reads
\begin{eqnarray}
S_n^-(\lambda)&=&S^-(\lambda)+\frac{F^0-S^0+n}{\lambda-a},\nonumber\\
S_n^0(\lambda)&=&S^0(\lambda)+\frac{F^0-S^0+n+b}{\lambda-a},\nonumber\\
S_n^-(\lambda)&=&S^-(\lambda)+\frac{F^0-S^0+n+2b}{\lambda-a},
\label{5.2}
\end{eqnarray}
where $a$ and $b$ are arbitrary complex
parameters and $F^0=\lim_{\lambda\rightarrow\infty}\lambda
F^0(\lambda)$.
It can be easily checked that operators (\ref{5.2})
actually satisfy the quasi-commutation relations (\ref{3.1}).

\medskip
The representation space of this algebra coincides by
construction with the representation space of the Gaudin algebra.
The lowest weight vector $|0\rangle$ is the same as in
section \ref{3} which gives us the possibility of checking
formulas (\ref{3.4}) by using their Gaudin analogues (\ref{2.5}).
Taking into account that
\begin{eqnarray}
S^0|0\rangle=F^0|0\rangle
\label{5.3}
\end{eqnarray}
and using formulas (\ref{2.5}), we obtain:
\begin{eqnarray}
S_n^0(\lambda)|0\rangle=\left(F^0(\lambda)+
\frac{b+n}{\lambda-a}\right)|0\rangle,\quad
S_n^-(\lambda)|0\rangle=\left(\frac{n}{\lambda-a}\right)|0\rangle.
\label{5.4}
\end{eqnarray}
We see that the lowest weight vector is actually not
annihilated by ``lowering'' operators if $n\neq 0$. 
Taking $n=0$ in (\ref{5.4}) we
obtain the expression for the lowest weight of quasi-Gaudin
algebra
\begin{eqnarray}
F(\lambda)=F^0(\lambda)+\frac{b}{\lambda-a}.
\label{5.5}
\end{eqnarray}
Now it is absolutely clear that algebra (\ref{3.1}) is a
continuous {\it deformation} of the Gaudin algebra. The role of the
deformation parameter is played by $a$. If
$a\rightarrow\infty$, then the $n$-dependence of the
operators (\ref{5.2}) disappears and they transform into
the ordinary generators of Gaudin algebra. Respectively, the
quasi-commutators in (\ref{3.1}) become the ordinary ones. 
As to the formulas (\ref{3.4}) defining the representations
of algebra (\ref{3.1}), they, in the limit $a\rightarrow\infty$,
also transform into relations (\ref{2.5}) for the Gaudin algebra.

\section{Quasi-exact solvability of quasi-Gaudin models}
\label{6}

Substituting (\ref{5.2}) into (\ref{3.2}), we obtain 
the following explicit form of 
quasi-Gaudin hamiltonians:
\begin{eqnarray}
H_n(\lambda)&=&S^0(\lambda)S^0(\lambda)-\frac{1}{2}
S^-(\lambda)S^+(\lambda)-\frac{1}{2}S^+(\lambda)S^-(\lambda)
+\nonumber\\
&+&\frac{2S^0(\lambda)(n+b+F^0-S^0)-S^-(\lambda)(n+2b+F^0-S^0)
-S^+(\lambda)(n+F^0-S^0)}{\lambda-a}+\nonumber\\
&+&\frac{b(b-1)}{(\lambda-a)^2}
\label{6.1}
\end{eqnarray}
Now we can check in 
an independent way that the operators (\ref{6.1})
actually describe quasi-exactly solvable models. 
Denote by $\Phi_n$ a linear hull of all vectors 
$S^+(\xi_k)\ldots S^+(\xi_1)|0\rangle$ with arbitrary
$\xi_1,\ldots,\xi_k$ and $k\le n$. It is known that
$\dim\Phi_n<\infty$ for any $n$ if $F^0(\lambda)$ is a rational 
function. From the obvious relations
$S^\pm(\lambda)\Phi_n\subset\Phi_{n\pm 1}$,
$S^0(\lambda)\Phi_n\subset \Phi_n$, 
$(n+F^0-S^0)\Phi_n \subset \Phi_{n-1}$
and $(n+F^0-S^0)\Phi_m\subset \Phi_m$ for $m\neq n$
it immediately follows that each of the operators $H_n(\lambda)$
admits only one invariant subspace, $\Phi_n$. This subspace
is finite-dimensional and therefore the models (\ref{6.1})
are quasi-exactly solvable in the standard sense 
of this word (for the ``standard'' quasi-exact solvability 
see e.g. refs. \cite{Sh89,MPRST90,Sh94}).

\section{Quasi-$sl(2)$ algebra}
\label{7}

It is known that the $sl(2)$ algebra can be considered as a
limiting case of the Gaudin ${\cal G}[sl(2)]$ algebra (\ref{2.1}). 
Its generators $S^0$, $S^-$ and $S^+$ are given by  
\begin{eqnarray}
S^0=\lim_{\lambda\rightarrow\infty}\lambda S^0(\lambda),\quad
S^-=\lim_{\lambda\rightarrow\infty}\lambda S^-(\lambda),\quad
S^+=\lim_{\lambda\rightarrow\infty}\lambda S^+(\lambda),
\label{7.1}
\end{eqnarray}
and satisfy the usual commutation relations
\begin{eqnarray}
S^0S^--S^-S^0&=&-S^-, \nonumber\\
S^0S^+-S^+S^0&=&+S^+, \nonumber\\
S^-S^+-S^+S^-&=&2S^0.
\label{7.2}
\end{eqnarray}
It is natural to ask what kind of algebra will appear
if we consider an analogous limit of quasi-${\cal G}[sl(2)]$ algebra? 
By analogy with the Gaudin case, we can define the 
generators of this algebra by the formulas
\begin{eqnarray}
S_n^0=\lim_{\lambda\rightarrow\infty}\lambda S_n^0(\lambda),\quad
S_n^-=\lim_{\lambda\rightarrow\infty}\lambda S_n^-(\lambda),\quad
S_n^+=\lim_{\lambda\rightarrow\infty}\lambda S_n^+(\lambda).
\label{7.3}
\end{eqnarray}
Multiplying the quasi-commutation relations (\ref{3.1}) by $\lambda\mu$
and tending both $\lambda$ and $\mu$ to infinity we easily
derive the relations between these generators, which read
\begin{eqnarray}
S_{n-1}^0S_n^--S_n^- S_n^0&=& -S_n^-, \nonumber\\
S_{n+1}^0S_n^+-S_n^+ S_n^0&=& +S_n^+, \nonumber\\
S_{n+1}^-S_n^+ -S_{n-1}^+S_n^-&=&2S_n^0.
\label{7.4}
\end{eqnarray}
We can consider (\ref{7.4}) as the definig relations of
a certain modification of the $sl(2)$ algebra which can be called
``quasi-$sl(2)$ algebra''. 
Despite the fact that it {\it is not} a Lie algebra, it has
many properties similar to those of the ordinary $sl(2)$.
For example, it has a quasi-analogue of the Casimir
operator,
\begin{eqnarray}
C_n=S_n^0S_n^0-\frac{1}{2}S_{n+1}^-S_n^+-
\frac{1}{2}S_{n-1}^+S_n^-
\label{7.5}
\end{eqnarray}
which quasi-commutes with all generators:
\begin{eqnarray}
C_nS_{n-1}^+=S_{n-1}^+C_{n-1},\quad
C_nS_{n+1}^-=S_{n+1}^-C_{n+1},\quad
C_nS_n^0=S_n^0C_n.
\label{7.6}
\end{eqnarray}
The ``representations'' of quasi-$sl(2)$ algebra can be
constructed in the same way as in the quasi-Gaudin case.
Defining the ``lowest weight vector'' $|0\rangle$ by 
\begin{eqnarray}
S_0^0|0\rangle=F|0\rangle,\qquad
S_0^-|0\rangle=0,
\label{7.7}
\end{eqnarray}
where $F$ is a certain arbitrarily fixed number,
we can define the ``representation space'' $W_{F}$
as a linear hull of vectors 
\begin{eqnarray}
|m\rangle_k=S_{m+k}^+S_{m-1+k}^+
\ldots S_{2+k}^+ S_{1+k}^+|0\rangle,
\label{7.8}
\end{eqnarray}
with arbitrary $k\in Z$ and $m\in N$.  If we are interested in the
spectrum of quasi-Casimir operator, then it is easily seen
that, in contrast with the standard $sl(2)$ case, it is
not infinitely-degenerate and 
contains only one exactly constructable eigenvector.
For example, the only possibility for vector (\ref{7.8}) to be an 
eigenvector of (\ref{7.5}) is realized when $k=-1$ and $m=n$. In this
case, the corresponding eigenvalue is equal to $F(F-1)$.
So, one can say that the quasi-Casimir operators represent
the simplest (and, in some sense, trivial) quasi-exactly
solvable models.

\medskip
It is remarkable that the generators of quasi-$sl(2)$
algebra can be realized in terms of the generators of ordinary $sl(2)$
algebra. The corresponding formulas can be obtained after
substituting formulas (\ref{5.2}) into (\ref{7.3}) and read
\begin{eqnarray}
S_n^-=S^-+n+F^0-S^0,\quad
S_n^0=n+b+F^0,\quad
S_n^+=S^++n+2b+F^0-S^0.
\label{7.9}
\end{eqnarray}
This realization corresponds to the choice $F=F^0+b$ in (\ref{7.7}).

\section{The Yangian and the $XXX$ model}
\label{8}

Let us give some brief exposition of facts concerning the
Yangian ${\cal Y}[sl(2)]$, its representations and properties of 
the so-called $XXX$ models associated with it.
A more detailed exposition of this subject can be found in
refs. \cite{Dr88,KoBoIz93}. The commutation relations for generators
$T_{\alpha\beta}(\lambda),\ \alpha,\beta=1,2$ can be
extracted from formulas (\ref{1.1}) and (\ref{1.6}).
Introducing the commonly used notations
\begin{eqnarray}
T_{11}(\lambda)=A(\lambda),\quad T_{12}(\lambda)=B(\lambda),\quad
T_{21}(\lambda)=C(\lambda),\quad T_{22}(\lambda)=D(\lambda),
\label{8.1}
\end{eqnarray}
we can rewrite these relations in more explicit form
\begin{eqnarray}
(\lambda-\mu)\{A(\lambda)B(\mu)-B(\mu)A(\lambda)\}&=&
\eta\{A(\mu)B(\lambda)-A(\lambda)B(\mu)\}\nonumber\\
(\lambda-\mu)\{A(\lambda)C(\mu)-C(\mu)A(\lambda)\}&=&
\eta\{A(\lambda)C(\mu)-A(\mu)C(\lambda)\}\nonumber\\
(\lambda-\mu)\{D(\lambda)B(\mu)-B(\mu)D(\lambda)\}&=&
\eta\{D(\lambda)B(\mu)-D(\mu)B(\lambda)\}\nonumber\\
(\lambda-\mu)\{D(\lambda)C(\mu)-C(\mu)D(\lambda)\}&=&
\eta\{D(\mu)C(\lambda)-D(\lambda)C(\mu)\}\nonumber\\
(\lambda-\mu)\{A(\lambda)D(\mu)-D(\mu)A(\lambda)\}&=&
\eta\{C(\mu)B(\lambda)-C(\lambda)B(\mu)\}\nonumber\\
(\lambda-\mu)\{B(\lambda)C(\mu)-C(\mu)D(\lambda)\}&=&
\eta\{D(\mu)A(\lambda)-D(\lambda)A(\mu)\}.
\label{8.2}
\end{eqnarray}
Using these relations one can easily prove that the operators
\begin{eqnarray}
H(\lambda)=A(\lambda)+D(\lambda)
\label{8.3}
\end{eqnarray}
commute with each other,
\begin{eqnarray}
[H(\lambda),H(\mu)]=0,
\label{8.4}
\end{eqnarray} 
for any $\lambda$ and $\mu$. This enables one to consider them
as integrals of motion of a completely integrable quantum system
which is known under name of $XXX$ 
model\footnote{More precisely --- the 
$XXX$ inhomogeneous magnetic chain.}.
The hamiltonians $H(\lambda)$ of these models act in the
representation space of algebra ${\cal Y}[sl(2)]$ which
is defined as a linear hull of vectors 
\begin{eqnarray}
|\xi_1,\ldots,\xi_n\rangle=C(\xi_n)\ldots C(\xi_1)|0\rangle
\label{8.5}
\end{eqnarray}
with arbitrary $n$ and $\xi_1,\ldots,\xi_n$.
Here $|0\rangle$ is the lowest weight vector, satisfying
the conditions
\begin{eqnarray}
B(\lambda)|0\rangle=0,\quad
A(\lambda)|0\rangle=a(\lambda)|0\rangle, \quad
D(\lambda)|0\rangle=d(\lambda)|0\rangle,
\label{8.6}
\end{eqnarray}
and functions $a(\lambda)$ 
and $b(\lambda)$ play the role of lowest
weight functions 
determining the form of the representation\footnote{Some 
formulas of this section differ a little bit from those used in the 
standerd literature 
since, instead of the standardly used highest weight representation
we are considering here the lowest weight one. But the results are,
of course, the same.}.
Denoting the representation space defined by formulas
(\ref{8.5}) and (\ref{8.6}) 
by $W_{a(\lambda),d(\lambda)}$, we can write down the 
spectral problem for $H(\lambda)$
\begin{eqnarray}
H(\lambda)\phi=E(\lambda)\phi,\quad \phi\in W_{a(\lambda),d(\lambda)}.
\label{8.7}
\end{eqnarray}
It turns out that this 
problem is exactly solvable within the Bethe ansatz
\begin{eqnarray}
\phi=C(\xi_n)\ldots C(\xi_1)|0\rangle.
\label{8.8}
\end{eqnarray}
This can be demonstrated exactly as in the Gaudin case. We
simply should act on (\ref{8.8}) by operator $H(\lambda)$
and transfer the operators $A(\lambda)$ and $D(\lambda)$
(from which it consists) to the right by using bilinear relations
\begin{eqnarray}
A(\lambda)C(\xi_i)=\frac{-\eta}{\lambda-\xi_i}C(\lambda)A(\xi_i)+
\frac{\lambda-\xi_i+\eta}{\lambda-\xi_i}C(\xi_i)A(\lambda),\nonumber\\
D(\lambda)C(\xi_i)=\frac{+\eta}{\lambda-\xi_i}C(\lambda)D(\xi_i)+
\frac{\lambda-\xi_i-\eta}{\lambda-\xi_i}C(\xi_i)D(\lambda),
\label{8.9}
\end{eqnarray}
which trivially follow from (\ref{8.2}). As soon as
operators $A(\lambda), D(\lambda)$ (as well as operators 
$A(\xi_i), D(\xi_i), \ i=1,\ldots, n$ 
arising after these transferences)
appear in front of the lowest weight vector $|0\rangle$,
they should be absorbed by it. The Bethe ansatz equations
guaranteeing the absence of the unwanted terms have the form
\begin{eqnarray}
\prod_{k=1, k\neq i}^n\frac{\xi_i-\xi_k-\eta}{\xi_i-\xi_k+\eta}
=\frac{a(\xi_i)}{d(\xi_i)},
\quad i=1,\ldots, n. 
\label{8.10}
\end{eqnarray}
and the final expression for the spectrum of the $XXX$ model reads
\begin{eqnarray}
E(\lambda)=
a(\lambda)\prod_{i=1}^n\frac{\lambda-\xi_i+\eta}{\lambda-\xi_i}
+d(\lambda)\prod_{i=1}^n\frac{\lambda-\xi_i-\eta}{\lambda-\xi_i}.
\label{8.11}
\end{eqnarray}
If, for example, $a(\lambda)$ and $d(\lambda)$ are rational
functions, then the number of solutions of Bethe ansatz equations 
(\ref{8.10}) is finite for any $n$. But since $n$ is an
{\it arbitrary} non-negative integer, the complete number of
solutions of problem (\ref{8.7}) is infinite. It can be
shown that these solutions expire all possible solutions of
problem (\ref{8.7}) and therefore it is exactly solvable.

\section{Quasi-Yangian and quasi-$XXX$ models}
\label{9}

Let us now construct the quasi-Yangian. Its generators 
depend additionally on an integer $n$ and
can be denoted by $A_n(\lambda)$, $B_n(\lambda)$, 
$C_n(\lambda)$ and 
$D_n(\lambda)$. The
quasi-commutation relations for these generators read
\begin{eqnarray}
(\lambda-\mu)\{A_{n-1}(\lambda)B_n(\mu)-B_n(\mu)A_n(\lambda)\}&=&
\eta\{A_{n-1}(\mu)B_n(\lambda)-A_{n-1}(\lambda)B_n(\mu)\}\nonumber\\
(\lambda-\mu)\{A_{n+1}(\lambda)C_n(\mu)-C_n(\mu)A_n(\lambda)\}&=&
\eta\{A_{n+1}(\lambda)C_n(\mu)-A_{n+1}(\mu)C_n(\lambda)\}\nonumber\\
(\lambda-\mu)\{D_{n-1}(\lambda)B_n(\mu)-B_n(\mu)D_n(\lambda)\}&=&
\eta\{D_{n-1}(\lambda)B_n(\mu)-D_{n-1}(\mu)B_n(\lambda)\}\nonumber\\
(\lambda-\mu)\{D_{n+1}(\lambda)C_n(\mu)-C_n(\mu)D_n(\lambda)\}&=&
\eta\{D_{n+1}(\mu)C_n(\lambda)-D_{n+1}(\lambda)C_n(\mu)\}\nonumber\\
(\lambda-\mu)\{A_n(\lambda)D_n(\mu)-D_n(\mu)A_n(\lambda)\}&=&
\eta\{C_{n-1}(\mu)B_n(\lambda)-C_{n-1}(\lambda)B_n(\mu)\}\nonumber\\
(\lambda-\mu)\{B_{n+1}(\lambda)C_n(\mu)-C_{n-1}(\mu)B_n(\lambda)\}&=&
\eta\{D_n(\mu)A_n(\lambda)-D_n(\lambda)A_n(\mu)\}.
\label{9.1}
\end{eqnarray}
It is not difficult to check that these relations are  invariant
under ``quasi-similarity'' transformations 
\begin{eqnarray}
A_n(\lambda)&\rightarrow&U_nA_n(\lambda)U_n^{-1},\nonumber\\
B_n(\lambda)&\rightarrow&U_{n-1}B_n(\lambda)U_n^{-1},\nonumber\\
C_n(\lambda)&\rightarrow&U_{n+1}C_n(\lambda)U_n^{-1},\nonumber\\
D_n(\lambda)&\rightarrow&U_nD_n(\lambda)U_n^{-1},
\label{9.1a}
\end{eqnarray}
in which $U_n$ are arbitrary invertible operators.

\medskip
Using (\ref{9.1}) one can easily prove that the operators
\begin{eqnarray}
H_n(\lambda)=A_n(\lambda)+D_n(\lambda)
\label{9.2}
\end{eqnarray}
commute with each other,
\begin{eqnarray}
[H_n(\lambda),H_n(\mu)]=0,
\label{9.3}
\end{eqnarray} 
for any $\lambda$ and $\mu$ and any given $n$. However, the
$H$-operators with different indices $n$ do not generally commute 
with each other. This leads us to a conclusion that we
deal with an infinite series of 
{\it different} completely integrable quantum systems which we
shall call the quasi-$XXX$-models.

\medskip
In order to 
formulate a spectral problem for hamiltonians $H_n(\lambda)$ 
one should construct a carrier space in which they act. 
This space, as usually, can be identified 
with a ``representation 
space'' of quasi-${\cal Y}[sl(2)]$ algebra. For constructing
the latter we can use the same reasonings as in section \ref{3} 
and define it as a linear hull of vectors 
\begin{eqnarray}
|\xi_1,\ldots,\xi_m\rangle_k=C_{m+k}(\xi_m)C_{m+k-1}(\xi_{m-1})
\ldots C_{k+2}(\xi_2)C_{k+1}(\xi_1)|0\rangle
\label{9.4}
\end{eqnarray}
with arbitrary $m\in N$, 
$k\in Z$ and $\xi_1,\ldots,\xi_m$. It is not difficult to see that,
due to commutation relations 
(\ref{9.1}), the vectors (\ref{9.4}) are symmetric 
with respect to all permutations of numbers 
$\xi_1,\ldots,\xi_m$.
Here, as before, 
$|0\rangle$ is the lowest weight vector, satisfying
the restricted conditions
\begin{eqnarray}
B_0(\lambda)|0\rangle=0,\quad
A_0(\lambda)|0\rangle=a(\lambda)|0\rangle, \quad
D_0(\lambda)|0\rangle=d(\lambda)|0\rangle,
\label{9.5}
\end{eqnarray}
and functions $a(\lambda)$ and $d(\lambda)$ play the role
of the lowest weight functions determining the form of the 
``representation''.
We denote such a ``representation space'' 
by $W_{a(\lambda),d(\lambda)}$.
In the next section we demonstrate 
that the lowest weight vector is actually not
annihilated by other lowering operators $B_n(\lambda)$ with
$n\neq 0$. Moreover, 
we show that it is even not an eigenvector of operators
$D_n(\lambda)$ if $n\neq 0$. In conclusion we present 
the final form of the spectral 
problem for operators $H_n(\lambda)$, which reads
\begin{eqnarray}
H_n(\lambda)\phi=E(\lambda)\phi, 
\quad \phi\in W_{a(\lambda),d(\lambda)}
\label{9.6}
\end{eqnarray}
and whose solutions will be discussed in the next section.

\section{Bethe ansatz for quasi-$XXX$ model}
\label{10}

The results of section \ref{9} enable 
one to show that spectral problems (\ref{9.6})
can be algebraically solved within the Bethe ansatz
\begin{eqnarray}
\phi=C_{m+k}(\xi_m)C_{m+k-1}(\xi_{m-1})
\ldots C_{k+2}(\xi_2)C_{k+1}(\xi_1)|0\rangle
\label{10.1}
\end{eqnarray}
only if the numbers $m\in N$ and $k\in Z$ 
are chosen as $m=n$ and $k=-1$,
where $n$ is the 
index of the hamiltonian. In this case (\ref{10.1}) takes
the form
\begin{eqnarray}
\phi=C_{n-1}(\xi_n)C_{n-2}(\xi_{n-1})
\ldots C_{1}(\xi_2)C_{0}(\xi_1)|0\rangle,
\label{10.2}
\end{eqnarray}
and we can use the ``quasi'' analogues of relations (\ref{8.9})
\begin{eqnarray}
A_n(\lambda)C_{n-1}(\xi_n)=\frac{-\eta}{\lambda-\xi_n}
C_{n-1}(\lambda)A_{n-1}(\xi_n)+
\frac{\lambda-\xi_n+\eta}{\lambda-\xi_n}C_{n-1}(\xi_n)
A_{n-1}(\lambda),\nonumber\\
D_n(\lambda)C_{n-1}(\xi_n)=\frac{+\eta}{\lambda-\xi_n}
C_{n-1}(\lambda)D_{n-1}(\xi_n)+
\frac{\lambda-\xi_n-\eta}
{\lambda-\xi_n}C_{n-1}(\xi_n)D_{n-1}(\lambda)
\label{10.3}
\end{eqnarray}
(which can be easily obtained 
from basic relations (\ref{9.1})) to transfer 
the operators $A_n(\lambda)$ and $D_n(\lambda)$
to the right. Exactly as in the case of quasi-Gaudin
algebra, each permutation of $A$- or $D$-operators with
$C$-operators forming Bethe vector decreases the indices of
the former by one. Finally, when all permutations will be
completed and the operators $A_n(\lambda), D_n(\lambda)$ 
together with the daughter operators 
$A_n(\xi_i), D_n(\xi_i), \ \xi=1,\ldots, n$ appear in front
of the lowest weight vector $|0\rangle$, they will have index $0$
and therefore can be 
absorbed by this vector by means of formulas (\ref{9.5}).
The condition of the cancellation of unwanted terms leads
to the sysem of Bethe ansatz equations
\begin{eqnarray}
\prod_{k=1, k\neq i}^n\frac{\xi_i-\xi_k-\eta}
{\xi_i-\xi_k+\eta}=\frac{a(\xi_i)}{d(\xi_i)},
\quad i=1,\ldots, n,
\label{10.4}
\end{eqnarray}
whose form exactly coincides with (\ref{8.10}),
and the corresponding eigenvalues read
\begin{eqnarray}
E(\lambda)=
a(\lambda)\prod_{i=1}^n\frac{\lambda-\xi_i+\eta}{\lambda-\xi_i}
+d(\lambda)\prod_{i=1}^n\frac{\lambda-\xi_i-\eta}{\lambda-\xi_i},
\label{10.5}
\end{eqnarray}
which is again exactly the same expression as in (\ref{8.11}). 
Summarising, we can say, that the use of quasi-${\cal Y}[sl(2)]$
algebra enables one to construct an infinite set of completely
integrable systems, each of which is only partially
solvable and has only a certain finite number of solutions 
(of course, provided that both $a(\lambda)$ and $b(\lambda)$ are 
rational functions). The second interesting fact is that the
we obtained exactly the same set of solutions as in the
case of ordinary Yangian. But now these solutions are
distributed between different models.

\section{The existence of quasi-Yangian}
\label{11}

In this section we present one of possible realizations
of quasi-Yangian. In full accordance with the Gaudin case, 
we construct it from generators of the
ordinary Yangian. Assume that the latter is 
characterized by four generators $A(\lambda), B(\lambda), 
C(\lambda), D(\lambda)$ 
acting in a certain representation space with lowest weights
$a^0(\lambda)$ and $d^0(\lambda)$. Introduce the limiting operator
\begin{eqnarray}
S^0=\frac{1}{2\eta}\lim_{\lambda\rightarrow\infty}\lambda(A(\lambda)-
D(\lambda))
\label{11.1}
\end{eqnarray}
and its lowest weight
\begin{eqnarray}
F^0=\frac{1}{2\eta}
\lim_{\lambda\rightarrow\infty}\lambda(a^0(\lambda)-
d^0(\lambda)).
\label{11.2}
\end{eqnarray}
Then the needed realization reads
\begin{eqnarray}
A_n(\lambda)=
A(\lambda)\left(1+\eta\frac{F^0-S^0+n-1/2+b}{\lambda-a}\right)+
B(\lambda)
\left(\eta\frac{F^0-S^0+n+2b}{\lambda-a}\right),\nonumber\\
B_n(\lambda)=
B(\lambda)\left(1-\eta\frac{F^0-S^0+n+1/2+b}{\lambda-a}\right)-
A(\lambda)
\left(\eta\frac{F^0-S^0+n}{\lambda-a}\right),\nonumber\\
C_n(\lambda)=
C(\lambda)\left(1+\eta\frac{F^0-S^0+n-1/2+b}{\lambda-a}\right)+
D(\lambda)
\left(\eta\frac{F^0-S^0+n+2b}{\lambda-a}\right),\nonumber\\
D_n(\lambda)=
D(\lambda)\left(1-\eta\frac{F^0-S^0+n+1/2+b}{\lambda-a}\right)-
C(\lambda)\left(\eta\frac{F^0-S^0+n}{\lambda-a}\right).
\label{11.3}
\end{eqnarray}
One can easily check by direct calculations that these operators
actually satisfy the quasi-commutation relations (\ref{9.1}) and
thus form the quasi-${\cal Y}[sl(2)]$.
Since the representation space of this quasi-Yangian coincides by
construction with the representation space of the ordinary Yangian,
we can check formulas (\ref{9.5}) by using 
their Yangian analogues (\ref{8.6}).
Taking into account that
\begin{eqnarray}
S^0|0\rangle=F^0|0\rangle,
\label{11.4}
\end{eqnarray}
and using formulas (\ref{8.6}), we obtain:
\begin{eqnarray}
A_n(\lambda)|0\rangle &=& a^0(\lambda)
\left(1+\eta\frac{n-1/2+b}{\lambda-a}\right)|0\rangle,
\nonumber\\
D_n(\lambda)|0\rangle &=& d^0(\lambda)
\left(1-\eta\frac{n+1/2+b}{\lambda-a}\right)|0\rangle-
\left(\eta\frac{n}{\lambda-a}\right)C(\lambda)|0\rangle,\nonumber\\
B_n(\lambda)|0\rangle &=&-a^0(\lambda)
\left(\eta\frac{n}{\lambda-a}\right)|0\rangle.
\label{11.5}
\end{eqnarray}
We see that the lowest weight vector is actually
annihilated by ``lowering'' operators $B_n(\lambda)$ and
is an eigenvalue of operators $D_n(\lambda)$ only for $n=0$.
Taking $n=0$ in (\ref{11.5}) we
obtain the expressions for the lowest weights of quasi-Yangian
\begin{eqnarray}
a(\lambda)=
a^0(\lambda)\left(1+\eta\frac{b-1/2}{\lambda-a}\right),\quad
d(\lambda)=
d^0(\lambda)\left(1-\eta\frac{b+1/2}{\lambda-a}\right).
\label{11.6}
\end{eqnarray}
Now it is absolutely clear that algebra (\ref{9.1}) is a
continuous deformation of the Yangian (\ref{8.2}). The role of the
deformation parameter is played by $a$. If
$a\rightarrow\infty$, then the $n$-dependence of the
operators (\ref{11.3}) disappears and they transform into
the ordinary generators of Yangian. Respectively, the
quasi-relations in (\ref{9.1}) become the ordinary ones. 
As to the formulas (\ref{9.5}) defining the representations
of algebra (\ref{9.1}), they, in the limit $a\rightarrow\infty$
also transform into the relations (\ref{8.6}) for the Yangian.

\section{Quasi-exact solvability of quasi-$XXX$ models}
\label{12}

Substituting (\ref{11.3}) into (\ref{9.2}), we obtain an explicit
form of hamiltonians of quasi-$XXX$ models:
\begin{eqnarray}
H_n(\lambda)&=&\frac{\lambda-a-\eta}
{\lambda-a}(A(\lambda)+D(\lambda))
\nonumber\\
&+&\frac{\eta}{\lambda-a}
(A(\lambda)-D(\lambda))(F^0-S^0+n+b)+\nonumber\\
&+&\frac{\eta}{\lambda-a}B(\lambda)(F^0-S^0+n+2b)-\nonumber\\
&-&\frac{\eta}{\lambda-a}C(\lambda)(F^0-S^0+n).
\label{12.1}
\end{eqnarray}
Let us now check in an 
independent way that the operators (\ref{12.1})
actually describe quasi-exactly solvable models. 
Denote by $\Phi_n$ the linear hull of all vectors 
$C(\xi_k)\ldots C(\xi_1)|0\rangle$ with arbitrary
$\xi_1,\ldots,\xi_k$ and $k\le n$. It is known that
$\dim\Phi_n<\infty$ for any $n$ if $a^0(\lambda)$ and
$d^0(\lambda)$ are rational functions.
From the obvious relations
$B(\lambda)\Phi_n\subset\Phi_{n-1}$, 
$C(\lambda)\Phi_n\subset\Phi_{n+1}$,
$A(\lambda)\Phi_n\subset \Phi_n$,
$D(\lambda)\Phi_n\subset \Phi_n$,
$(F^0-S^0+n)\Phi_n \subset \Phi_{n-1}$
and $(F^0-S^0+n)\Phi_m\subset \Phi_m$ for $m\neq n$
it immediately follows that each of the operators $H_n(\lambda)$
admits only one algebraically constructable 
invariant subspace, $\Phi_n$. This subspace
is finite-dimensional and therefore the models (\ref{12.1})
are quasi-exactly solvable.

\section{Historical roots}
\label{13}

Quasi-exactly solvable models came to the light about ten years ago.
After first examples which have been found 
heuristically \cite{ZaUl84, BaVs86, TuUs87},
the three general theories explaining the phenomenon of
quasi-exact solvability and presenting constructive ways
for building such models have been formulated.
These theories are known under names of ``partial algebraization 
method'' \cite{Tu88,Sh89,MPRST90,GKO92,GKO94,Tu94,Sh94,GKO96}, 
``the inverse method of separation of 
variables'' \cite{Us88b,Us89,Us94} 
and ``the projection method'' \cite{Us92,Us94}. 
In sections \ref{14} -- \ref{16} we try to interpret the 
results obtained in previous sections from the point of view of
these three methods and 
show that, in contrast with the first method, which 
is not applicable to the study 
of quasi-${\cal T}_R$ algebras at all, 
the last two ones enable one to reproduce some general features of the 
proposed formalizm. Especially, this relates to the third,
projection method, by means of which we essentially obtained all
the results exposed in sections \ref{2} -- \ref{12}.

\section{The partial algebraization method}
\label{14}

One of the first theories of quasi-exactly solvable systems
proposed in 1988 in ref. \cite{Tu88} and called the ``partial
algebraization method'' was based on a purely algebraic construction. 
Due to simplicity
of this theory, its essence can be formulated in several words.

\subsection*{The method} 

The main idea of the partial algebraization method was based
on the observation that any finite-dimensional
representation of any Lie or Lie$_q$ algebra can be realized by
means of first order differential or 
pseudo-differential operators acting in the
spaces of functions. Let $L_i,\ i=1,\ldots, d$ be such generators.
Taking their combinations of the form
\begin{eqnarray}
H=\sum_{i,k=1}^d C_{ik}L_i L_k + \sum_{i=1}^d C_i L_i
\label{14.1}
\end{eqnarray}
one obtains a certain second-order 
differential or pseudo-differential operator
having (by construction) a finite-dimensional invariant subspace.
This means that for this operator at least a certain finite part 
of the spectrum can be 
constructed algebraically (for more detail see refs. 
\cite{Tu88,Sh89,MPRST90,GKO92,GKO94,Tu94,Sh94,GKO96}). 
Note also that this approach 
is easily extendable to the matrix case 
(see e.g. refs. \cite{Sh89,BrKo94}).

\subsection*{Discussion} 

The partial algebraization method seems to be rather general.
The models obtainable in such a way are not-neccessarily
integrable because only an existence of algebraically constructable
finite-dimensional invariant subspaces is required.
But just because of the generality of this method, 
it contains an essential deficiency
not giving a constructive way of solving the resulting
algebraic problems. Essentially, the method stops as soon
as the representatibility of a certain interesting
hamiltonian in the form (\ref{14.1}) is proven. 

What can we say of quasi-Gaudin and quasi-$XXX$ spin models
from the point of view of this theory? Unfortunately, nothing,
because neither the generators of the initial 
${\cal T}_R$ algebra used 
in formulas (\ref{2.3}) and (\ref{8.3}) 
nor their quasi-couterparts used in (\ref{3.2}) and
(\ref{9.2}) are assumed to 
realize any finite-dimensional representation.
The problem of finding algebras and their
finite-dimensional representations which would allow one to represent
our hamiltonians in the form (\ref{14.1}) is 
interesting and
we hope that it will find its solution in the nearest future.

\section{The inverse method of separation of variables}
\label{15}

Another oldest theory of quasi-exactly solvable systems  
proposed in 
1988 in ref. \cite{Us88b} and developed in \cite{Us89,Us94} 
was purely analytic.
It can be divided into 
two parts: 1) construction of one-dimensional exactly-
and quasi-exactly solvable multi-parameter spectral equations
and 2) reduction of these 
equations to ordinary one-parameter exactly- or quasi-exactly 
solvable spectral equations in a multi-dimensional space.
Consider this method for the simplest case of second-order equations.

\subsection*{The first part of the method}

Let $\partial$ denote the differential operator, 
$\partial\phi(\lambda)=\phi'(\lambda)$, and let $F(\lambda)$ be
a given rational function. Consider the equation
\begin{eqnarray}
(\partial-F(\lambda))^2\Psi(\lambda)=E(\lambda)\Psi(\lambda)
\label{15.1}
\end{eqnarray}
for two functions $E(\lambda)$ and $\Psi(\lambda)$. It can
be shown that it admits a very simple class of solutions
\begin{eqnarray}
\Psi(\lambda)=\prod_{i=1}^n(\lambda-\xi_i),
\quad n=0,1,2,\ldots,
\label{15.2}
\end{eqnarray}
\begin{eqnarray}
E(\lambda)=F'(\lambda)+F^2(\lambda)+
2\sum_{i=1}^n\frac{F(\lambda)-F(\xi_i)}
{\lambda-\xi_i},\quad n=0,1,2,\ldots,
\label{15.3}
\end{eqnarray}
where, for any given $n$, the 
numbers $\xi_1,\ldots,\xi_n$ satisfy the system of
numerical equations
\begin{eqnarray}
\sum_{k=1,k\neq i}^n\frac{1}{\xi_i-\xi_k}+
F(\xi_i)=0,\quad i=1,\ldots,n.
\label{15.4}
\end{eqnarray}
Before discussing the mathematical meaning of the obtained
solutions (which obviously form only a very small subset in
the set of all possible solutions of equation (\ref{15.1})), 
let us consider an analogous 
construction for pseudo-differential operators.
Let $T_\eta$ denote the shift operator, 
$T_\eta\phi(\lambda)=\phi(\lambda+\eta)$, 
and $a(\lambda), d(\lambda)$
be some rational functions on $\lambda$. Consider the equation
\begin{eqnarray}
(a(\lambda)T_\eta+
d(\lambda)T_\eta^{-1})\Psi(\lambda)=E(\lambda)\Psi(\lambda)
\label{15.5}
\end{eqnarray}
for two functions $E(\lambda)$ and $\Psi(\lambda)$. It can
be shown that it admits a very simple class of solutions
\begin{eqnarray}
\Psi(\lambda)=\prod_{i=1}^n(\lambda-\xi_i),
\quad n=0,1,2,\ldots,
\label{15.6}
\end{eqnarray}
\begin{eqnarray}
E(\lambda)=
a(\lambda)\prod_{i=1}^n\frac{\lambda-\xi_i+\eta}{\lambda-\xi_i}
+d(\lambda)\prod_{i=1}^n\frac{\lambda-\xi_i-\eta}{\lambda-\xi_i},
\quad n=0,1,2,\ldots,
\label{15.7}
\end{eqnarray}
where, for any given $n$, 
the numbers $\xi_1,\ldots,\xi_n$ satisfy the system of
numerical equations
\begin{eqnarray}
\prod_{k=1, k\neq i}^n\frac{\xi_i-\xi_k-\eta}
{\xi_i-\xi_k+\eta}=\frac{a(\xi_i)}{d(\xi_i)},
\quad i=1,\ldots, n,
\label{15.8}
\end{eqnarray}
An analogous construction 
for pseudo-differential equations of a little bit different type 
was considered recently in ref. \cite{WiZa95}.

\medskip
Now let us explain the reason for which these solutions are
interesting to us. A remarkable fact (distinguishing
these solutions from an infinite number of other solutions) is that
for any rational function $F(\lambda)$ (in differential
case) and for any pair of rational functions $a(\lambda)$ 
and $d(\lambda)$
(in pseudo-differential case) 
the function $E(\lambda)$ is representable in the form
\begin{eqnarray}
E(\lambda)=W_0(\lambda)+\sum_{\alpha=1}^N E_\alpha W_\alpha(\lambda),
\label{15.9}
\end{eqnarray} 
where $N$ is some number depending only on the form of 
function $F(\lambda)$
or, respectively, of $a(\lambda)$ and $d(\lambda)$,
and $W_n(\lambda)$ are fixed functions not depending on $\xi_i$. 
All the dependence of $E(\lambda)$ on numbers
$\xi_i$ satisfying the equations (\ref{15.4}) or (\ref{15.8}) 
is concentrated in the numerical coefficients $E_\alpha$.
It can be shown that for any given $n$ both equation
(\ref{15.4}) and (\ref{15.8}) have exactly $(N-1+n)!/((N-1)!)n!)$ 
solutions.
Since $n$ in formulas (\ref{15.2}), (\ref{15.3}) and
(\ref{15.6}), (\ref{15.7}) can be made arbitrarily large, 
a total number of algebraic solutions of equations
(\ref{15.1}) and (\ref{15.5}) is infinite.
These reasonings enable one to interpret (\ref{15.1}) 
and (\ref{15.5}) as 
a certain exactly 
(algebraically) solvable $N$-parameter spectral equation.
Indeed, introducing the notation
\begin{eqnarray}
D(\lambda)=\left\{
\begin{array}{ll}
(\partial-F(\lambda))^2-W_0(\lambda),& \mbox{in
differential case,}\\
a(\lambda)T_\eta+d(\lambda)T_\eta^{-1}-W_0(\lambda),&
\mbox{in pseudo-differential case,}
\end{array}
\right.
\label{15.10}
\end{eqnarray}
we can rewrite the equations (\ref{15.1}) and (\ref{15.5}) in
the following common form
\begin{eqnarray}
D(\lambda)\Psi(\lambda)= 
\left(\sum_{\alpha=1}^N E_\alpha 
W_\alpha(\lambda)\right)\Psi(\lambda).
\label{15.11}
\end{eqnarray}
In particular case, 
when $N=1$, (\ref{15.11}) reduces to an ordinary
one-parameter spectral equation which, in turn, can easily be
reduced to the Schr\"odinger form. This gives us the list
of known one-dimensional exactly solvable models.

A remarkable feature of 
equation (\ref{15.11}) is that the spectrum
of one of its spectral parameters is always degenerate,
i.e. depends only on $n$ but not on the numbers $\xi_i$.
Assume for definiteness that the parameter $E_N$ has such a
degenerate spectrum. 
Then, introducing the notation $E_N=E_{n,N}$ and taking
\begin{eqnarray}
D_n(\lambda)=\left\{
\begin{array}{ll}
(\partial-F(\lambda))^2-
W_0(\lambda)-E_{n,N}W_N(\lambda),& \mbox{in
differential case,}\\
a(\lambda)T_\eta+d(\lambda)
T_\eta^{-1}-W_0(\lambda)-E_{n,N}W_N(\lambda),&
\mbox{in pseudo-differential case,}
\end{array}
\right.
\label{15.12}
\end{eqnarray}
we can represent (\ref{15.11}) as an infinite sequence of
$(N-1)$-parameter spectral equations
\begin{eqnarray}
D_n(\lambda)\Psi(\lambda)= \left(\sum_{\alpha=1}^{N-1}
E_\alpha W_\alpha(\lambda)\right)
\Psi(\lambda),\quad n=0,1,2,\ldots
\label{15.13}
\end{eqnarray}
each of which will obviously be only quasi-exactly solvable.
For $N=2$ these equations can again be
considered as one-parameter spectral equations easily
reducible to the Schr\"odinger form. This gives us the list
of all known series of one-dimensional quasi-exactly solvable models.
This completes the first part of the method.

\subsection*{The second part of the method} 

Now, what to do if $N>1$ and $N>2$ in the first and second
cases, respectively? Remember that any one-dimensional 
$d$-parameter spectral
equation with spectral parameters $E_1,\ldots,E_d$ 
can be interpreted as a result of separation of
variables in a certain 
$d$-dimensional separable (and hence integrable) system.
The parameters 
$E_1,\ldots, E_d$ play in this case the role of separation
constants and their admissible values have the meaning of 
the eigenvalues of some commuting $d$-dimensional operators
--- the integrals of motion of the abovementioned system.
Note that the form of these integrals can easily be reconstructed by
means of the so-called inverse procedure of separation of variables.

Let us first find 
the $N$ integrals of motion, $H_1,\ldots,H_N$ for the first
equation (\ref{15.11}) and, after this, construct their combination
\begin{eqnarray}
H(\lambda)=W_0(\lambda)+\sum_{\alpha=1}^N H_\alpha W_\alpha(\lambda).
\label{15.14}
\end{eqnarray} 
The operators $H(\lambda)$ will obviously commute with each
other, 
\begin{eqnarray}
[H(\lambda),H(\mu)]=0,
\label{15.15}
\end{eqnarray} 
and will have the eigenvalues 
given by formulas (\ref{15.3}) or (\ref{15.7}).
This means that the system described by these operators
will not be only integrable but also exactly solvable.

Let us now find the integrals of motion associated with the
second system of multi-parameter spectral equations (\ref{15.13}).
Since the form of these equations depends explicitly on
$n$, the obtained integrals also should depend on $n$.
Denoting them by $H_{n,1},\ldots, H_{n,N-1}$, let us
construct their analogous combinations
\begin{eqnarray}
H_n(\lambda)=W_0(\lambda)+ E_{n,N}W_N(\lambda)+\sum_{\alpha=1}^{N-1} 
H_{n,\alpha} W_\alpha(\lambda),\quad n=0,1,2,\ldots
\label{15.16}
\end{eqnarray} 
From this formula it immediately follows that operators $H_n(\lambda)$
form again commutative families
\begin{eqnarray}
[H_n(\lambda),H_n(\mu)]=0,
\label{15.17}
\end{eqnarray} 
for any $n$, but they are not obliged 
to commute with each other if their indices 
do not coincide. The eigenvalues of operators
$H_n(\lambda)$ will be described 
by the same formulas (\ref{15.3}) or (\ref{15.7})
but with fixed $n$. Therefore all the integrable models
represented by these operators 
will have only a finite number of explicitly constructible
solutions and thus will be quasi-exactly solvable.
This completes the second part of the method.

\subsection*{Discussion} 

Now what can we say of this method from
the point of view 
of results obtained in sections \ref{2} -- \ref{12}?
A little bit more than in the case of partial
algebraization method. First of all, 
we arrived to the same situation as in these sections.
The quasi-exactly solvable models have one and
the same set of solutions as the exactly solvable ones 
but these solutions are distributed between different models.
Second, the 
spectra (\ref{15.3}) (resp. (\ref{15.7})) of our models and
the form of auxiliary 
conditions (\ref{15.4}) (resp. (\ref{15.8})) for numbers $\xi_i$
exactly coincide 
with the spectra and Bethe ansatz equations for the
Gaudin (resp. $XXX$) and quasi-Gaudin 
(resp. quasi-$XXX$) models. This suggests that we
obtained the same models but in a very special coordinate form.

Summarizing, one can say that the inverse method of
separation of variables correctly reproduces some external
featurs of models discussed in sections \ref{2} -- \ref{12} .
However, and this is quite obvious, 
the analytic formalizm exposed in this section
is not convenient for investigating the properties of the
underlying ${\cal G}[sl(2)]$ (resp. ${\cal Y}[sl(2)]$) and 
quasi-${\cal G}[sl(2)]$ 
(resp. quasi-${\cal Y}[sl(2)]$) algebras. Moreover, it
is applicable only to the study of separable models
and thus cannot pretend on generality.

\section{The projection method}
\label{16}

The third method of building quasi-exactly solvable models
proposed in 1992 in ref. \cite{Us92} 
and developed in \cite{Us94} is the so-called projection method. 
This method is applicable to all integrable and 
exactly solvable systems with global symmetry.
The idea of this method is very simple and, in fact, it is
implicitly encoded in the inverse method of separation of
variables which we discussed in the previous section.
Indeed, as it follows from the above discussion, the 
complete separation
of variables in both $N$-dimensional exactly solvable 
and $(N-1)$-dimensional quasi-exactly solvable models
(constructable in the framework of the inverse method)
leads (by construction) to one and the same
multi-parameter spectral equation. But this suggests that
the $(N-1)$-dimensional quasi-exactly solvable models can be
obtained from their $N$-dimensional exactly solvable
conterparts by means of the partial separation of variables.
Why this observation is so important for us? Because it
gives us a possibility of formulating a new more general method for
constructing quasi-exactly solvable models. The generality
of this method follows from the fact that it does not
require anymore 
a complete separability of variables in the initial
exactly solvable 
model. The only thing which one needs indeed is the
requirement of a 
partial separability which, roughly speaking, can always be
satisfied if the hamiltonian of an initial exactly
solvable model admits some global Lie symmetry. 
Let us now explain in more detail what we mean.

\subsection*{The method} 

Let $H(\lambda)$ be some quantum hamiltonian
acting in a certain linear vector 
space $\Phi$, and let the spectral problem
\begin{eqnarray}
H(\lambda)\phi=E(\lambda)\phi,\quad \phi\in \Phi
\label{16.1}
\end{eqnarray}
have an infinite 
and discrete set of exactly constructable solutions.  
Assume that it is possible to
represent $H(\lambda)$ in the form of a differential or 
pseudo-differential 
operator in certain variables $x_1,\ldots, x_N$
and consider $\Phi$ as a space of polynomials in $x_1,\ldots,x_N$.
Obviously, for most of interesting systems this is
possible.

\medskip
Assume now that there exists some operator $L$ which: a) 
commutes with $H(\lambda)$:
\begin{eqnarray}
[H(\lambda), L]=0,
\label{16.2}
\end{eqnarray}
b) is representable in the form of a first-order
differential operator in the same variables $x_1,\ldots, x_N$:
\begin{eqnarray}
L=\sum_{i=1}^N A_i(x_1,\ldots,x_N)\frac{\partial}{\partial x_i}+
B(x_1,\ldots,x_N)
\label{16.3}
\end{eqnarray}
and c) has an infinite and discrete spectrum in $\Phi$.
Again, for most of interesting systems such operator does
actually exist.

\medskip
Let $l_n,\ n=0,1,2\ldots$ and $\Phi_n,\ n=0,1,2,\ldots$ denote the
eigenvalues and corresponding eigensubspaces of the operator
$L$ in $\Phi$. 
It follows from (\ref{16.2}) that all these eigensubspaces, 
which, by definition consist of polynomials in $x_1,\ldots,x_N$, 
are simultaneously
the invariant subspaces for $H(\lambda)$.
This means that all spectral problems
\begin{eqnarray}
H(\lambda)\phi=E(\lambda)\phi,\quad \phi\in \Phi_n,\quad n=0,1,2,\ldots
\label{16.4}
\end{eqnarray}
are well defined and are exactly solvable.

\medskip
Let us now denote by $\Psi_n$ the sets of all 
(not neccessarily normalizable)
solutions of the equation
\begin{eqnarray}
L\psi=l_n\psi, \quad n=0,1,2,\ldots.
\label{16.5}
\end{eqnarray}
Due to the absence of the condition of polynomiality
of these solutions, these sets
should be wider than $\Phi_n$:
\begin{eqnarray}
\Phi_n\subset\Psi_n, \quad n=0,1,2,\ldots,
\label{16.6}
\end{eqnarray}
but, as before, they will 
be the invariant subspaces for $H(\lambda)$,
and threfore the extended versions of equations (\ref{16.4})
\begin{eqnarray}
H(\lambda)\psi=
E(\lambda)\psi,\quad \psi\in \Psi_n,\quad n=0,1,2,\ldots
\label{16.7}
\end{eqnarray}
will also be well defined and, formally,
will be quasi-exactly solvable. This, however,
will not be that kind of quasi-exact solvability which we
discussed before and which we want to have. 
This will be rather some pathologically
trivial case of it, because the
hamiltonian $H(\lambda)$ still does know that it should act
only in one of the sets $\Psi_n$ and 
acts equally well in all of them.
Practically this means that we cannot interpret the
equations (\ref{16.7}) as spectral equations for {\it different} 
models.

\medskip
In order to improve the situation, one should 
restrict somehow the action of the hamiltonian $H(\lambda)$ 
to one of the sets $\Psi_n$.
But for this it is 
neccesary to have a little bit more information of
the structure of these sets. We can extract it from the
equation (\ref{16.5}) which, according to (\ref{16.3}), 
can be rewritten as
\begin{eqnarray}
\left(\sum_{i=1}^N 
A_i(x_1,\ldots,x_N)\frac{\partial}{\partial x_i}\right)
\psi(x_1,\ldots,x_N)=\biggl(l_n-B(x_1,\ldots,x_N)\biggr)
\psi(x_1,\ldots,x_N).
\label{16.8}
\end{eqnarray}
The general solution of this equation has the form 
\begin{eqnarray}
\psi(x_1,\ldots,x_N)=\rho_n(x_1,\ldots,x_n)f[\sigma_1(x_1,\ldots,x_n),
\ldots,\sigma_{N-1}(x_1,\ldots,x_N)]
\label{16.9}
\end{eqnarray}
where $\rho_n$ is a certain arbitrarily fixed 
particular solution
of (\ref{16.8}) and $\sigma_i,\ i=1,\ldots,
N-1$ are some functionally independent solutions of
the auxiliary equation
\begin{eqnarray}
\left(\sum_{i=1}^N 
A_i(x_1,\ldots,x_N)\frac{\partial}{\partial x_i}\right)
\sigma(x_1,\ldots,x_N)=0.
\label{16.10}
\end{eqnarray}
The function $f(\sigma_1,\ldots,\sigma_{N-1})$ in
(\ref{16.9}) is arbitrary and therefore all the information
of the specific properties of the spaces $\Psi_n$ is
concentrated in $\rho_n$. The set of all functions 
$f(\sigma_1,\ldots,\sigma_{N-1})$, which we denote by $F$,
contains for any given $n$ a certain subset $F_n$ corresponding
to explicit solutions of the initial equations (\ref{16.4}).

\medskip
Let us now note that, due to the functional independence of
$N$ functions
\begin{eqnarray}
\rho&=&[\rho_n(x_1,\ldots,x_n)]^{\frac{1}{n}},\nonumber\\
\sigma_i&=&\sigma_1(x_1,\ldots,x_n),\quad i=1,\ldots, N-1,
\label{16.11}
\end{eqnarray}
they, for any given $n$, can be considered as the new
independent variables (instead of $x_1,\ldots,x_N$).
Rewriting the hamiltonian $H(\lambda)$ 
in terms of these new variables and 
using the obtained 
expressions (\ref{16.9}) for functions $\psi$ with
given $n$, we can rewrite equations (\ref{16.7}) in the form:
\begin{eqnarray}
\tilde H_n(\lambda)f(\sigma_1,\ldots,\sigma_{N-1})=
E(\lambda)f(\sigma_1,\ldots,\sigma_{N-1}),
\quad f \in F,\quad n=0,1,2,\ldots
\label{16.12}
\end{eqnarray}
where
\begin{eqnarray}
\tilde H_n(\lambda)=\rho_n^{-1}H(\lambda)\rho_n,\quad n=0,1,2,\ldots
\label{16.13}
\end{eqnarray}

\medskip
Now the only one last step remains. The step which just
suggested us to call this method the projection method.
Let us look at formula (\ref{16.12}). We see that the transformed
hamiltonians $\tilde H_n(\lambda)$ are, by construction,
certain differential or pseudo-differential operator in
{\it all}
variables $\rho$ and $\sigma_1,\ldots,\sigma_{N-1}$.
At the same time, they act only on functions on $N-1$ variables 
$\sigma_1,\ldots,\sigma_{N-1}$
and give again the functions of the same variables.
This means that the most general form of these operators is
\begin{eqnarray}
\tilde H_n(\lambda)=H_n(\lambda)+H_n'(\lambda)\frac{\partial}{\partial
\rho},\quad n=0,1,2,\ldots,
\label{16.14}
\end{eqnarray}
where $H_n'(\lambda)$ are certain operators in all
variables $\rho_n$ and $\sigma_1,\ldots,\sigma_N$, and $H_n(\lambda)$
are operators in variables $\sigma_1,\ldots,\sigma_N$ only. 
This enables one to rewrite (\ref{16.12}) in the following
final form
\begin{eqnarray}
H_n(\lambda)f(\sigma_1,\ldots,\sigma_{N-1})=
E(\lambda)f(\sigma_1,\ldots,\sigma_{N-1}),
\quad f \in F,\quad n=0,1,2,\ldots
\label{16.15}
\end{eqnarray}
We see that these equations have the form of ordinary
quasi-exactly solvable ones, because the set $F$ contains
by construction the subsets $F_n$ in which these
equations are exactly solvable. All the information of a
quasi-exactly solvable model is encoded now in its hamiltonian
which explicitly depends on $n$.

\subsection*{Discussion} 

Let us now try to explain the results of
sections \ref{2} --- \ref{12} from the point of view 
of the projection method. For this, let us assume that 
$H(\lambda)$ is a hamiltonian of Gaudin or $XXX$ model 
expressed respectively via generators of ${\cal G}[sl(2)]$ 
or ${\cal Y}[sl(2)]$ algebras by formulas (\ref{2.3}) or (\ref{8.3}). 
It is known that generators of both ${\cal G}[sl(2)]$ and ${\cal
Y}[sl(2)]$ algebras can be constructed from generators 
$S^-_\alpha$, $S^0_\alpha$, $S^+_\alpha$, $\alpha=1,\ldots,N$ of
algebra $sl(2)\otimes\ldots\otimes sl(2)$ 
($N$ times). The corresponding
formulas have the form
\begin{eqnarray}
\left(
\begin{array}{cc}
S^0(\lambda) & -S^-(\lambda) \\
+S^+(\lambda) & -S^0(\lambda)
\end{array}
\right)
=\sum_{\alpha=1}^N
\left(
\begin{array}{cc}
S^0_\alpha & -S^-_\alpha \\
+S^+_\alpha & -S^0_\alpha
\end{array}
\right)
\frac{1}{\lambda-a_\alpha},
\label{16.a}
\end{eqnarray}
for Gaudin algebra ${\cal G}[sl(2)]$, and
\begin{eqnarray}
\left(
\begin{array}{cc}
A(\lambda) & B(\lambda) \\
C(\lambda) & D(\lambda)
\end{array}
\right)
=\prod_{\alpha=1}^N
\left[ 1 + \frac{\eta}{\lambda-a_\alpha}\left(
\begin{array}{cc}
S^0_\alpha & -S^-_\alpha \\
+S^+_\alpha & -S^0_\alpha
\end{array}
\right)\right],
\label{16.b}
\end{eqnarray}
for the Yangian ${\cal Y}[sl(2)]$. Using formulas
(\ref{2.3}) and (\ref{8.3}), we can respectively express the
hamiltonians $H(\lambda)$ of both Gaudin and $XXX$ models
via generators $S^-_\alpha$, $S^0_\alpha$, $S^+_\alpha$, 
$\alpha=1,\ldots,N$.
Remember that these generators can be realized as 
differential operators
\begin{eqnarray}
S_\alpha^-=\frac{\partial}{\partial x_\alpha},\quad
S_\alpha^0=x_\alpha\frac{\partial}{\partial x_\alpha}+F_\alpha,\quad
S_\alpha^+=x_\alpha^2\frac{\partial}
{\partial x_\alpha}+2F_\alpha x_\alpha
\label{16.16}
\end{eqnarray}
acting on polynomials in variables $x_1,\ldots,x_N$.
Formulas (\ref{16.a}) and (\ref{16.b}) together with (\ref{16.16})
enable one to represent the hamiltonian $H(\lambda)$ in
the form of some differential operator in $N$ variables
$x_1,\ldots,x_N$. The space $\Phi$ in which this hamiltonian acts
becomes respectively the space of polynomials.

\medskip
Now note that, 
the hamiltonian $H(\lambda)$ has a global $sl(2)$ symmetry, 
which, in particular, implies its commutativity with the
operator of the $z$-projection of total spin
\begin{eqnarray}
L=\sum_{\alpha=1}^N S_\alpha^0.
\label{16.17}
\end{eqnarray}
According to (\ref{16.16}), the differential form of this
operator is 
\begin{eqnarray}
L=\sum_{\alpha=1}^N x_\alpha\frac{\partial}{\partial x_\alpha}+F,
\label{16.18}
\end{eqnarray}
where $F=F_1+\ldots+F_N$.
The eigenvalues of this operator in $\Phi$
are given by $l_n=n+F$ and the 
corresponding eigensubspaces $\Phi_n$ are
formed by all homogeneous polynomials of degree $N$.

\medskip
To construct the spaces $\Psi_n$, one should find a
general solution of the equation
\begin{eqnarray}
\left(\sum_{\alpha=1}^N 
x_\alpha\frac{\partial}{\partial x_\alpha}+F\right)
\psi(x_1,\ldots,x_N)=(n+F)\psi(x_1,\ldots,x_N)
\label{16.19}
\end{eqnarray}
which, obviously, reads
\begin{eqnarray}
\psi(x_1,\ldots,x_N)=x_N^n f\left(\frac{x_1}{x_N},\ldots,
\frac{x_{N-1}}{x_N}\right).
\label{16.20}
\end{eqnarray}
The new variables can hence be chosen in the form: 
\begin{eqnarray}
\rho=x_N,\quad \sigma_i=\frac{x_i}{x_N},\quad i=1,\ldots,N-1.
\label{16.21}
\end{eqnarray}

\medskip
According to 
general prescriptions of the projection method, the first
(preliminary) step 
in the reduction procedure lies in rewriting the
hamiltonian $H(\lambda)$ in new variables (\ref{16.21}) and in 
constructing the homogeneously transformed hamiltonians
$\tilde H_n(\lambda)$, according to formula (\ref{16.13}).
In our special case this formula gives:
\begin{eqnarray}
\tilde H_n(\lambda)=\rho^{-n}H(\lambda)\rho^n.
\label{16.g}
\end{eqnarray}
Using formulas (\ref{2.3}) and (\ref{8.3}), we obtain for
$\tilde H_n(\lambda)$:
\begin{eqnarray}
\tilde H_n(\lambda)=\tilde S_n^0(\lambda)
\tilde S_n^0(\lambda)-\frac{1}{2}
\tilde S_n^-(\lambda)\tilde S_n^+(\lambda)-
\frac{1}{2}\tilde S_n^+(\lambda)\tilde S_n^-(\lambda),
\label{16.h}
\end{eqnarray}
\begin{eqnarray}
\tilde S_n^0(\lambda)=\rho^{-n}S^0(\lambda)\rho^n,\quad
\tilde S_n^-(\lambda)=\rho^{-n}S^-(\lambda)\rho^n,\quad
\tilde S_n^+(\lambda)=\rho^{-n}S^+(\lambda)\rho^n,
\label{16.i}
\end{eqnarray}
for the Gaudin case, and
\begin{eqnarray}
\tilde H_n(\lambda)=\tilde A_n(\lambda)+\tilde D_n(\lambda),
\label{16.k}
\end{eqnarray}
\begin{eqnarray}
\tilde A_n(\lambda)=\rho^{-n}A(\lambda)\rho^n,\quad
\tilde D_n(\lambda)=\rho^{-n}D(\lambda)\rho^n,
\label{16.l}
\end{eqnarray}
for the case of Yangian. An explicit computation of
operators staying in the right hand sides of formulas
(\ref{16.h}) and (\ref{16.k}) gives
\begin{eqnarray}
\left(
\begin{array}{cc}
\tilde S_n^0(\lambda) & -\tilde S_n^-(\lambda) \\[0.1cm]
+\tilde S_n^+(\lambda) & -\tilde S_n^0(\lambda)
\end{array}
\right)
=
\left(
\begin{array}{cc}
\bar S^0(\lambda) & -\rho^{-1}\bar S^-(\lambda) \\[0.1cm]
+\rho \bar S^+(\lambda) & -\bar S^0(\lambda)
\end{array}
\right)
+
\nonumber\\[0.2cm]
+
\left(
\begin{array}{ll}
\rho\frac{\partial}{\partial\rho}+F_N+n-\sigma & 
-\frac{\partial}{\partial\rho}-\frac{n-\sigma}{\rho}  \\[0.1cm]
\rho^2\frac{\partial}{\partial\rho}+2F_N\rho+(n-\sigma)\rho & 
-\rho\frac{\partial}{\partial\rho}-F_N-(n-\sigma)
\end{array}
\right)
\frac{1}{\lambda-a_N},
\label{16.c}
\end{eqnarray}
for Gaudin algebra ${\cal G}[sl(2)]$, and
\begin{eqnarray}
\left(
\begin{array}{cc}
\tilde A_n(\lambda) & \tilde B_n(\lambda) \\[0.1cm]
\tilde C_n(\lambda) & \tilde D_n(\lambda)
\end{array}
\right)
=
\left(
\begin{array}{cc}
\bar A(\lambda) & \rho^{-1}\bar B(\lambda) \\[0.1cm]
\rho \bar C(\lambda) & \bar D(\lambda)
\end{array}
\right)
\times
\nonumber\\[0.2cm]
\times
\left[ 1 + \frac{\eta}{\lambda-a_N}\left(
\begin{array}{ll}
\rho\frac{\partial}{\partial\rho}+F_N+(n-\sigma) & 
-\frac{\partial}{\partial\rho}-\frac{n-\sigma}{\rho}  \\[0.1cm]
\rho^2\frac{\partial}{\partial\rho}+2F_N\rho+(n-\sigma)\rho & 
-\rho\frac{\partial}{\partial\rho}-F_N-(n-\sigma)
\end{array}
\right)\right],
\label{16.e}
\end{eqnarray}
for the Yangian ${\cal Y}[sl(2)]$. Here 
\begin{eqnarray}
\sigma=\sum_{\alpha=1}^{N-1}
\sigma_\alpha\frac{\partial}{\partial\sigma_\alpha}
\label{16.aa}
\end{eqnarray}
and the entries of the matrices 
stayng in the right hand sides of (\ref{16.c})
and (\ref{16.e}) are given by
\begin{eqnarray}
\left(
\begin{array}{cc}
\bar S^0(\lambda) & -\bar S^-(\lambda) \\[0.1cm]
+\bar S^+(\lambda) & -\bar S^0(\lambda)
\end{array}
\right)
=\sum_{\alpha=1}^{N-1}
\left(
\begin{array}{ll}
\sigma_\alpha\frac{\partial}{\partial\sigma_\alpha}+F_\alpha & 
-\frac{\partial}{\partial\sigma_\alpha}  \\[0.1cm]
\sigma_\alpha^2\frac{\partial}{\partial\sigma_\alpha}+
2F_\alpha\sigma_\alpha & 
-\sigma_\alpha\frac{\partial}{\partial\sigma_\alpha}-F_\alpha
\end{array}
\right)
\frac{1}{\lambda-a_\alpha}\qquad
\label{16.d}
\end{eqnarray}
and
\begin{eqnarray}
\left(
\begin{array}{cc}
\bar A(\lambda) & \bar B(\lambda) \\[0.1cm]
\bar C(\lambda) & \bar D(\lambda)
\end{array}
\right)
=\prod_{\alpha=1}^{N-1}
\left[ 1 + \frac{\eta}{\lambda-a_\alpha}\left(
\begin{array}{ll}
\sigma_\alpha\frac{\partial}{\partial\sigma_\alpha}+F_\alpha & 
-\frac{\partial}{\partial\sigma_\alpha}  \\[0.1cm]
\sigma_\alpha^2\frac{\partial}{\partial\sigma_\alpha}+
2F_\alpha\sigma_\alpha & 
-\sigma_\alpha\frac{\partial}{\partial\sigma_\alpha}-F_\alpha
\end{array}
\right)\right].
\label{16.f}
\end{eqnarray}

\medskip
In order to complete the reduction of the initial
hamiltonian $H(\lambda)$ to a sequence of hamiltonians $H_n(\lambda)$
of quasi-exactly solvable models, one should project the operators 
(\ref{16.h}) and (\ref{16.k}) onto a space of functions 
depending on variables $\sigma_1,\ldots,\sigma_{N-1}$ only. 
After some simple algebra one obtains
\begin{eqnarray}
H_n(\lambda)=S_n^0(\lambda)S_n^0(\lambda)-\frac{1}{2}
S_{n+1}^-(\lambda)S_n^+(\lambda)-
\frac{1}{2}S_{n-1}^+(\lambda)S_n^-(\lambda)
\label{16.m}
\end{eqnarray}
with
\begin{eqnarray}
S_n^-(\lambda)&=&
\bar S^-(\lambda)+\frac{n-\sigma}{\lambda-a_N},\nonumber\\
S_n^0(\lambda)&=&
\bar S^0(\lambda)+\frac{n-\sigma+F_N}{\lambda-a_N},\nonumber\\
S_n^-(\lambda)&=&
\bar S^-(\lambda)+\frac{n-\sigma+2F_N}{\lambda-a_N},
\label{16.mm}
\end{eqnarray}
for the Gaudin case, and
\begin{eqnarray}
H_n(\lambda)=A_n(\lambda)+D_n(\lambda)
\label{16.n}
\end{eqnarray}
with
\begin{eqnarray}
A_n(\lambda)=\bar A(\lambda)\left(1+\eta\frac{n-\sigma-1/2+F_N}
{\lambda-a_N}\right)+
\bar B(\lambda)
\left(\eta\frac{n-\sigma+2F_N}{\lambda-a_N}\right),\nonumber\\
B_n(\lambda)=\bar B(\lambda)\left(1-\eta\frac{n-\sigma+1/2+F_N}
{\lambda-a_N}\right)-
\bar A(\lambda)
\left(\eta\frac{n-\sigma}{\lambda-a_N}\right),\nonumber\\
C_n(\lambda)=\bar C(\lambda)\left(1+\eta\frac{n-\sigma-1/2+F_N}
{\lambda-a_N}\right)+
\bar D(\lambda)
\left(\eta\frac{n-\sigma+2F_N}{\lambda-a_N}\right),\nonumber\\
D_n(\lambda)=\bar D(\lambda)\left(1-\eta\frac{n-\sigma+1/2+F_N}
{\lambda-a_N}\right)-
\bar C(\lambda)\left(\eta\frac{n-\sigma}{\lambda-a_N}\right).
\label{16.nn}
\end{eqnarray}
for the case of Yangian. 

\medskip
It is not difficult to see that formulas (\ref{16.m}) and (\ref{16.n})
exactly coincide with expressions (\ref{3.2}) and (\ref{9.2})
for hamiltonians of quasi-Gaudin and quasi-$XXX$ models. As
to the formulas (\ref{16.mm}) and (\ref{16.nn}), they, up
to a trivial change of notations, are identical to formulas
(\ref{5.2}) and (\ref{11.3}) for generators of ${\cal G}[sl(2)]$
and ${\cal Y}[sl(2)]$ algebras.

\medskip
Summarizing, we can say that the projection method enables one
to reproduce correctly the structure of quasi-Gaudin and quasi-$XXX$
models together with all properties of the undelying ${\cal G}[sl(2)]$
and ${\cal Y}[sl(2)]$ algebras. However, this method
contains so many technicalities and is so much model dependent 
that its practical usefulness seems rather doubtful. 
In next sections I will expose
another simple method which practically without any
calculations allows one to construct quasi-analogue for any
${\cal T}_R$ algebra.

\section{Graded associative algebras}
\label{17}

Now we start the exposition of a general and model independent 
method for building quasi-${\cal T}_R$ 
algebras and their ``representations''.
The idea of this method is based on the observation that both
${\cal T}_R$ and quasi-${\cal T}_R$ 
algebras can be considered as two {\it different}
realizations of one and the same abstract 
$Z^r$-graded\footnote{Here $Z^r=Z\otimes\ldots\otimes Z$ ($r$ times) 
denotes the set of all $r$-dimensional
vectors with integer components.} unital associative algebra $A$. 
The difference between the ``ordinary'' and ``quasi'' versions 
of this algebra lies in a different nature of its elements 
and also in different 
definition of the corresponding associative product.
In this section we remind the reader some general properties of
an abstract algebra 
$A$ realized in an abstract linear vector space $V$. 
The structure of the pair $A, V$ is described by the following 
standard axioms. 

\medskip
{Axiom 1}. The set $A$ is a union of 
non-intersecting layers $A^n,\ n\in Z^r$ each of which 
forms a linear space over $C$. If $a\in A^n$ then we say that  
$a$ has grading $n$ and write this as $g(a)=n$. 
The summation of elements belonging to different layers is forbidden.

\medskip
{ Axiom 2}. For any $n, m\in Z^r$ there
exists a multiplication map $(A^n,A^m)\rightarrow A^{n+m}$ which puts 
into correspondence to any pair of 
elements $a\in A^n$ and $b\in A^m$ their product 
$ab\in A^{n+m}$. This product is a) associative: $a(bc)=(ab)c$, for any 
$a\in A^n$, $b\in A^m$, $c\in A^l$ and $n, m, l \in Z^r$,
b) distributive: $(a+b)c=ac+bc$, $c(a+b)=ca+cb$ for any
$a,b\in A^n$, $c\in A^m$ and $n,m\in Z^r$, c) has a unit $I\in A^0$ 
such that $Ia=aI=a$ for any $a\in A^n$ and $n\in Z^r$.

\medskip
{ Axiom 3}. The set $V$ is a union of 
non-intersecting layers $V^n,\ n\in Z^r$ each of which
forms a linear space over $C$. If $v\in V^n$ then we say that $v$ has
grading $n$ and write this 
as $g(v)=n$. The summation of elements belonging to
different layers is forbidden.

\medskip
{ Axiom 4}. For any $n,m\in Z^r$ there
exists a multiplication map $(A^n,V^m)\rightarrow V^{n+m}$ which puts 
into correspondence to any pair of elements
$a\in A^n$ and $v\in V^m$ their product $av\in
V^{n+m}$. This product is a) associative:
$a(bv)=(ab)v$, for any $a\in A^n$, $b\in A^m$,
$v\in V^l$ and $n, m, l \in Z^r$,
b) distributive: $(a+b)v=av+bv$, $a(v+u)=av+au$ for any
$a,b\in A^n$, $v,u\in V^m$ and $n,m\in Z^r$, c) has the same 
unit $I\in A_0$ such that $Iv=vI=v$ for
any $v\in V^n$ and $n\in Z^r$.

\medskip
Any $A$-algebra contains two important subalgebras. One,
which we denote by $A^0$, is formed by all zero-graded elements of $A$.
The second one which we denote by $A^+$ consists of all
elements with non-negative gradings. The non-negativity of
a grading means that all components of 
the grading vector $n=(n_1,\ldots,n_r)$
are non-negative integers.  The complement of $A^+$ in $A$,
consisting of elements whose gradings contain at least one
negative component (we call such gradings quasi-negative), 
does not form any subalgebra of $A$. Nevertheless, this complement 
plays an important role in applications and we shall denote it by $A^-$. 

\medskip
The space $V$ also contains an important subspace formed by
all vectors with nonnegative gradings. We denote it by $V^+$.
The importance of the space $V^+$ 
lies in the fact that it, under some special conditions,
takes the meaning of an 
algebraically constructable representation space of 
algebra $A$. In order to realize these conditions one needs a
special so-called lowest weight vector $|0\rangle$ with grading 
zero which is annihilated by any operator from $A^-$:
\begin{eqnarray}
a^-|0\rangle=0, \quad \mbox{for any}\quad a^-\in A^-
\label{17.1}
\end{eqnarray}
and is an eigenvector of all operators from $A^0$:
\begin{eqnarray}
a^0|0\rangle=\epsilon(a^0)|0\rangle,\quad
\mbox{for any}\quad a^0\in A^0.
\label{17.2}
\end{eqnarray}
Here $\epsilon(a^0)$ denotes some linear functional of $a^0$
which is usually called the lowest weight.
Once a vector $|0\rangle$ is constructed, one can construct
the space $V^+$ in a purely algebraic way, simply 
by acting on it by various elements of $A^+$:
\begin{eqnarray}
V^+=\bigcup_{a^+\in A^+} a^+|0\rangle.
\label{17.3}
\end{eqnarray}

\medskip
Note that the operator elements of $A$-algebras can sometimes be
treated as quantum observables and the vectors of the space
$V^+$ --- as physical states. In this case the
grading of vectors $v$ from $V^+$ can be considered as a collection 
of ``occupation numbers'' for certain elementary excitations
and the action of operators from $A^+$ and $A^-$ on these vectors
--- as a process of creation and annihilation of excitations.
The lowest weight vector takes respectively 
the meaning of the vaccuum state.
If the number of excitations is a 
conserved quantity, then the hamiltonian of a system
should neccessarily be a zero-graded element of subalgebra $A^0$.
Fixing the hamiltonian 
$h^0\in A^0$, the vaccuum state $|0\rangle\in V^0$ and
the space $V^+$ we essentially fix a physical system.

\medskip
Once a physical system is fixed, one can state the problem of
finding all its physical states. This leads us to an
eigenvalue problem for hamiltonian $h^0$:
\begin{eqnarray}
h^0\phi=\epsilon\phi, \quad \phi\in V^+.
\label{17.4}
\end{eqnarray}
Note that the knowledge of only general properties of algebra $A$ listed
above is, obviously, not sufficient to present an explicit solution
of problem (\ref{17.4}).In order to solve it, we need an information of 
more specific properties of this algebra which strongly depend on a
concrete form of its elements and therefore are not encoded in axioms 1
-- 4. In fact, what we actually need, is the existence of 
special bilinear relations between the elements of algebra $A$.

\medskip
Indeed, assume that for given $h^0\in A^0$ and each $m\ge 0$ there exists 
an element $a^+_m\in V^+$ with grading $m$ for which the relations 
\begin{eqnarray}
h^0\cdot a^+_m=a^+_m\cdot b^0_m+
\sum_{i}a^+_{m_i}\cdot a^-_{m_i}
\label{17.5}
\end{eqnarray}
hold. Here no special requirements to elements 
$b^0_m\in A^0$, $a^-_{m_i}\in A^-$
and $a^+_{m_i}\in A^+$ are assumed. Acting by this
bilinear operator relation on $|0\rangle$ and using formulas
(\ref{17.1}) and (\ref{17.2}), we obtain
\begin{eqnarray}
h^0\phi_m=\epsilon_m\phi_m,\quad  \epsilon_m=\epsilon(b^0_m),\quad
\phi_m=a^+_m|0\rangle, \quad m\in Z
\label{17.6}
\end{eqnarray}
which gives us infinitely many purely algebraic solutions of the 
spectral problem (\ref{17.4}).
Vectors $\phi_m=a^+_m|0\rangle$ solving this
problem are typical Bethe vectors.
Despite the seeming simplicity of this solution, its actual
construction is sometimes very complicated. The most
non-trivial thing is to find 
appropriate Bethe vectors for a given hamiltonian.
Fortunately, for many physically interesting systems this
problem has been successfully 
solved (see e.g. sections \ref{2} and \ref{8}).

\section{Modification of an associative product}
\label{18}

Up to now the consideration of $A$-algebras and their representations was
rather general: the elements of operator algebra $A$ were
considered as {\it elementary} objects not having any internal structure.
In this section we consider a special realization of algebra
$A$ whose elements are some {\it composite operators} and their product is
defined in a non-standard way. 
We demonstrate that such a realization also has
a lot of interesting physical 
applications. In order to distinguish this realization from
the general operator algebra $A$ discussed in section \ref{17} 
we shall use for it the notation $\hat A$ and reserve the star-symbol 
for the product of its elements. 

\medskip
We define the internal structure of the elements $\hat a\in
\hat A$ as follows. We consider them as functions on $Z^r$
whose particular values $a[n]$ are some linear operators on
$V$. The product of two elements $\hat a$ and
$\hat b$ will be denoted by $\hat a * \hat b$ and defined as
\begin{eqnarray}
(\hat a * \hat b)[n]=
a[n+g(\hat b)]\cdot b[n]
\label{18.1}
\end{eqnarray}
where the dot means the standard operator multiplication.
The action of an element $\hat a$ on the vector $v \in V$
is defined as
\begin{eqnarray}
\hat a  v=a[g(v)]v.
\label{18.2}
\end{eqnarray}
It is not difficult to check that these definitions do not 
contradict the axioms 1 -- 4.

\medskip
As in general case, we 
can introduce the subsets $\hat A^0$, $\hat A^+$ and 
$\hat A^-$ of algebra $\hat A$ consisting of the elements with zero, 
non-negative and quasi-negative gradings.
The values of these subsets on $Z^r$ we respectively denote by $A^0[n]$, 
$A^+[n]$ and $A^-[n]$. 

\medskip
In order to construct the 
lowest weight representation $V^+$ of algebra $\hat A$ we
need the lowest weight 
vector $|0\rangle\in V$ satisfying the hat-analogues of 
equations (\ref{17.1}) and (\ref{17.2}):
\begin{eqnarray}
\hat a^-|0\rangle=0, \quad \mbox{for any}\quad \hat a^-\in \hat A^-,
\label{17.1a}
\end{eqnarray}
\begin{eqnarray}
\hat a^0|0\rangle=\epsilon(\hat a^0)|0\rangle,\quad
\mbox{for any}\quad \hat a^0\in \hat A^0.
\label{17.2a}
\end{eqnarray}
If such a vector does exist, we can define the representation space as
\begin{eqnarray}
V^+=\bigcup_{\hat a^+\in \hat A^+} \hat a^+|0\rangle.
\label{17.3a}
\end{eqnarray}
Let us now rewrite formulas (\ref{17.1a}), (\ref{17.2a}) 
and (\ref{17.3a}) in components. Using definition (\ref{18.2}) and 
taking into account that the grading of the lowest
weight vector is zero, we obtain:
\begin{eqnarray}
a^-[0]|0\rangle=0, \quad \mbox{for any}\quad a^-[0]\in A^-[0],
\label{18.3}
\end{eqnarray}
\begin{eqnarray}
a^0[0]|0\rangle=\epsilon(a^0[0])|0\rangle,\quad
\mbox{for any}\quad a^0[0]\in A^0[0],
\label{18.4}
\end{eqnarray}
and
\begin{eqnarray}
V^+=\bigcup_{a^+[0]\in A^+[0]} a^+[0]|0\rangle.
\label{18.5}
\end{eqnarray}
From the above consideration it follows that 
the lowest weight vector $|0\rangle$ 
is not obliged to be annihilated by
all ``lowering'' operators 
$a^-[n]$ (with $n\neq 0$) and to be an eigenvector of all 
``neutral'' operators $a^0[n]$ (with $n\neq 0$).

\medskip
Why the $\hat A$-algebras maight be interesting to physicists?
There are several reasons for this. Let us try to explain one of them.
Assume that we managed to realize one and the same abstract $Z^r$-graded
unital associative algebra (abstract in the sense
that it is specified only by some relations between its
elements without any concretization of their internal structure)
in two different ways: 1) as an operator algebra $A$ and
2) as an algebra of composite operators $\hat A$, i.e. in terms of
the dot- and star-products, respectively.
Then we can say that algebras $A$ and $\hat A$ are isomorphic.

\medskip
Assume now that algebra $A$ 
contains two operators $a$ and $b$ commuting with each other:
\begin{eqnarray}
a\cdot b= b\cdot a.
\label{18.6}
\end{eqnarray}
If we consider this relation from the physical point of
view, we can imagine that the operators $a$ and $b$
represent two observables which can be measured simultaneously.
After this one can try to 
make some conclusions about their common spectral 
properties, etc. In other words, the relation (\ref{18.6}) 
may be quite meaningful and informative for physicists.

\medskip
Due to the assumed isomorphism between the $A$ 
and $\hat A$ algebras,
an analogous relation can also be written for 
the corresponding composite 
operators $\hat a$ and $\hat b$. 
They also should commute with each other:
\begin{eqnarray}
\hat a * \hat b= \hat b * \hat a.
\label{18.7}
\end{eqnarray}
But what can we now say of formula (\ref{18.7}) as physicists? 
At first sight, nothing, 
because the composite operators and their star products do not
have themselves 
any clear physical meaning. Remember, however, that the
composite operators have 
the meaning of the sets of ordinary operators and their
star product can be expressed via ordinary operator product. 
May be,
the relation (\ref{18.7}) 
will become more meaningful after rewriting it
in components? Let us see. Using formula (\ref{18.1}),
we can write:
\begin{eqnarray}
a[n+g(b)]\cdot b[n]=
b[n+g(a)]\cdot a[n]
\label{18.8}
\end{eqnarray}
Now the relation (\ref{18.8}) contains everything what we
need: the operators and operator products. However, we also
see that this formula does not describe any longer a
commutativity of any two objects: in general, this is a 
certain relation
between four different objects! Let us assume, however,
that both $a$ and $b$ elements have zero grading.
In this case the situation drastically changes: the
relation (\ref{18.8}) takes the form
\begin{eqnarray}
a[n]\cdot b[n]=
b[n]\cdot a[n]
\label{18.9}
\end{eqnarray}
and says us that the operators $a[n]$ and $b[n]$
commute in the standard operator sense for any $n$.
So, we see that we obtained even more than expected.
Instead of a single pair of commuting operators which we
started with, we obtained an infinite set of pairs of such operators!

\medskip
What does this mean from 
the physical point of view? There are many examples 
of $A$-algebras containing 
large families of mutually commuting operators. 
If the grading of these operators 
is zero, then it is possible to interpret them as
integrals of motion (hamiltonians) of some completely integrable
quantum systems. Assume now 
that we managed to construct a $\hat A$-analog 
of the above $A$-algebra 
which is isomorphic to $A$. Then we can repeat the
reasonings given above 
and, instead of a single family of commuting operators 
representing a single integrable system, construct an infinite and
discrete set of families of commuting operators which can
be treated as integrals of motion of infinitely many
different integrable quantum systems.  

\medskip
If we actually want to treat the infinite sets of commuting
operators appearing in our scheme as hamiltonians of some
physical systems, we should 
discuss the possible ways of solving the 
corresponding spectral problems.
Let us try to do this in 
the same way as in the case of general $A$-algebras
(see section \ref{17}). 
Let $h^0$ be a zero-graded element of algebra $A$ considered as
a hamiltonian of some integrable 
quantum system. Assume that spectral problem for this
hamiltonian is exactly solvable. This implies the existence of the
relations (\ref{17.5}) 
in algebra $A$. Assume also that we have in our 
disposal an algebra $\hat A$ 
isomorphic to $A$. Then, using this isomorphism, 
we can rewrite relations (\ref{17.5}) in the $\hat A$-algebraic form:
\begin{eqnarray}
\hat h^0* \hat a^+_m=\hat a^+_m*\hat b^0_m+
\sum_{i}\hat a^+_{m_i}*\hat a^-_{m_i}.
\label{18.10}
\end{eqnarray}
Here, as before, $m$ is an arbitrary non-negative element of $Z^r$ 
and no special requirements 
to the elements $\hat b^0_m\in \hat A^0$, $\hat a^-_{m_i}\in 
\hat A^-$ and $\hat a^+_{m_i}\in \hat A^+$ are assumed. 
Acting by (\ref{18.10}) on the 
lowest weight vector $|0\rangle$ and using formulas
(\ref{17.1a}) and (\ref{17.2a}) we obtain
\begin{eqnarray}
\hat h^0\phi_m= \epsilon(\hat b^0_m)
\phi_m, \quad \phi_m=\hat a^+_m|0\rangle.
\label{18.11}
\end{eqnarray}
Formula (\ref{18.11}) does not have yet any clear physical
meaning. In order to make it physicaly meaningful, one
should rewrite it in components. 
Taking into account that the grading of vector
$\phi_m$ is $m$ and using 
formula (\ref{18.2}), we obtain immediately
\begin{eqnarray}
h^0[m]\phi_m=\epsilon_m\phi_m,\quad 
\epsilon_m=\epsilon(b^0_m[0]),\quad
\phi_m=a^+_m[0]|0\rangle,\quad m\in Z^r
\label{18.13}
\end{eqnarray}
which gives us a purely 
algebraic solution of the spectral problems for 
$h^0[m], \ m\in Z^r$. Now however, this solution is 
incomplete, because the vectors
$\phi_m=a^+_m[0]|0\rangle$ 
with any given $m$ form only a negligibly
small part of the 
whole space $V^+$. This enables one to make a general 
conclusion that if the mother integrable system
associated with the $A$-algebra is exactly solvable,
i.e. admits a complete algebraic solution of the
spectral problem, then the daugther integrable systems
associated with the $\hat A$-algebra will be only
quasi-exactly solvable, i.e. only some finite parts of
their spectra can be constructed algebraically.

\section{Construction of modified graded associative algebras}
\label{19}

In this section we describe a simple scheme which essentially opens a 
practical way for building the graded algebras of composite operators. 
We demonstrate that any $A$-algebra whose elements admit
a special semi-differential realization can be reduced to a
certain $\hat A$-algebra which is isomorphic to $A$.

\medskip
Assume that each element $a$ of a given algebra $A$
(which is obviously an operator) admits the representation
\begin{eqnarray}
a=e^{t\cdot g(a)}(h(a)+h_i(a)\partial_i+
h_{ik}(a)\partial_i\partial_k+\ldots)
\label{19.1}
\end{eqnarray}
where $t=\{t_1,\ldots,t_r\}$ is a certain variable vector 
and $\partial_i=\partial/\partial t_i, \
i=1,\ldots,r$ are partial differential operators with respect to
the components of $t$.  The expression in the exponent is a scalar
product of two $r$-dimensional vectors $t$ and $g(a)$. 
The symbols $h(a), h_i(a), h_{ik}(a),\ldots$ 
denote some $t$-independent linear operators acting in a certain 
vector space $W$. A concrete form of these operators
depends on the element $a$. The summation over repeated 
indices is also assumed.
In this case, the elements of $A$ can be viewed as operators
in the space $W=V\times T$ where $T$ denotes the space of all
analytic functions of $t$. 
Assume also that any vector $w$ of the space $W$ admits the
representation
\begin{eqnarray}
w=e^{t\cdot g(w)}v(w)
\label{19.2}
\end{eqnarray}
where $v(w)$ denotes some $t$-independent vector of $V$.

\medskip
Let us now associate with elements $a$ defined by formula
(\ref{19.1}) some operator-valued functions $\hat a$ on $Z^r$ whose
particular values, $a[n]$, are operators in the space
$V$ defined by the following relations
\begin{eqnarray}
a[n]=h(a)+h_i(a) n_i+h_{ik}(a) n_i n_k+
\ldots
\label{19.3}
\end{eqnarray}
It is not difficult to check by means of direct computations that 
the elements $\hat a$ form an $\hat A$-algebra with the
products defined by formulas (\ref{18.1}) and (\ref{18.2}). This means
that any algebra $A$ whose elements admit the
representation (\ref{19.1}) and act in the space $V$ whose vectors
are representable form (\ref{19.2}) can be 
reduced to a certain $\hat A$ algebra. 

\medskip
The next important observation which can be immediately 
made after analysing formulas (\ref{19.1}),(\ref{19.2}) and (\ref{19.3})
is that the map $A\rightarrow \hat A$ we constructed is an isomorphism. 
This means that if there are some special (polynomial) relations 
between the elements of the operator algebra $A$ then the map
$A\rightarrow \hat A$ will induce the same relations 
between the corresponding
elements of $\hat A$. The only difference will concern the
definition of the product because the ordinary
operator product in $A$ should be replaced by a
rather specific product in $\hat A$.

\section{Quasi-Lie and quasi-Lie$_q$ algebras}
\label{20}

It is known that any Lie or Lie$_q$ algebra ${\cal L}$ of
the rank $r$ can be considered as a partially $Z^r$-graded algebra. 
The grading  can be introduced in many
different ways, but there is a special, the so-called root grading,
which seems to be most natural. In this case the gradings of
separate elements of a 
given algebra form a certain finite subset in $Z^r$ 
which describes the root system of an algebra 
in the basis of $r$ simple roots. 
This fact enables one to endow with grading also the
associated universal enveloping algebra ${\cal U}({\cal L})$.
This, obviously, will be a full $Z^r$-grading which makes it possible
to consider ${\cal U}({\cal L})$ as an $A$-algebra
described in section \ref{17}. 

\medskip
Now note that each algebra ${\cal L}$ can be realized in the form 
of differential or pseudo-differential operators in several variables
$x_1,\ldots,x_N$. A minimal 
possible number $N$ of these variables is $r$ 
(i.e. is equal to the rank of algebra ${\cal L}$ and to dimension of
grading vectors), and the maximal one is not limited. The
differential operators representing the elements of algebra
${\cal L}$ may depend additionally on certain numerical parameters
characterizing the representation  of algebra ${\cal L}$
and determining the eigenvalues of zero-graded elements of
${\cal L}$. Therefore, the grading of these parameters also should be
equal to zero. From this observation it automatically follows that 
the variables $x_1,\ldots,x_N$ should neccessarily be graded.
More exactly, at least $r$ of these variables should be graded and 
these gradings should be linearly independent. Otherwise,
it would be impossible to construct from them the
generators associated with simple roots whose gradings are
linearly independent by definition. 

\medskip
Taking into account the last reasonings, let us now try to reduce the 
generators $L^a$ of algebra ${\cal L}$ to a special canonical form 
by means of an appropriate change of variables. Note that
the existence of $r$ variables with linearly
independent gradings makes it possible 
to choose the new ones in such a way that to have $r$ 
variables $\rho_1,\rho_2,\ldots,\rho_r$ with gradings
$\{1,0,\ldots,0\}, \{0,1,\ldots,0\}, \ldots,
\{0,0,\ldots,1\}$ and all the remaining variables with grading zero.
The most general form of generators $L^a$ in the new variables reads
\begin{eqnarray}
L^a=\rho_1^{n^a_1}\cdots\rho_r^{n^a_r}\left(H^i+\sum_{i=1}^r H^a_i
\rho_i\frac{\partial}{\partial\rho_i}+\ldots\right)
\label{20.1}
\end{eqnarray}
where $n^a=(n^a_1,\ldots, n^a_r)$ is the gradings of $L^a$
and $H^a$, $H^a_i$, etc. are certain differential operators in
the remaining variables. The expression (\ref{20.1}) is,
however, rather cumbersome and therefore, it is convenient
to make one extra change of variables, namely, 
$\rho_i=e^{t_i},\ i=1,\ldots,r$,
which brings (\ref{20.1}) to the desired canonical form
\begin{eqnarray}
L^a=e^{{ t}{ g}(L^a)}(H^a+H^a_i\partial_i+\ldots)
\label{20.2}
\end{eqnarray}
with ${ t}=\{t_1,\ldots,t_r\}$. From this expression it
immediately follows that each element of the universal
enveloping algebra 
${\cal U}({\cal L})$ is also representable in the form
(\ref{19.1}) and for this reason ${\cal U}({\cal L})$ can be
reduced to a $\hat A$-algebra isomorphic to $A$.

\medskip
Let us now look at the explicit formulas.
Let ${\cal L}$ be a Lie algebra. Then the commutation
relations for its generators read
\begin{eqnarray}
L^a\cdot L^b-L^b\cdot L^a=C^{ab}_c\ L^c.
\label{20.3}
\end{eqnarray}
Rewritting the star-analogues of these relations (which
read similarly) in components $L^a[n]$ we obtain the relations
\begin{eqnarray}
L^a[n+ g(L^b)]\cdot L^b[{ n}]-
L^b[n+ g(L^a)]\cdot L^a[{ n}]=
C^{ab}_c\ L^c[{ n}]
\label{20.4}
\end{eqnarray}
which do not form any Lie algebra with respect to the
ordinary operator multiplication. We call such an algebra the 
quasi-Lie algebra and the expression staying in the left-hand
side of (\ref{20.4}) --- the quasi-commutator.
Let now ${\cal L}$ be a Lie$_q$ algebra. Then the most
general form of the relations for its generators is
\begin{eqnarray}
L^a\cdot L^b=R^{ab}_{cd}\ L^c\cdot L^d.
\label{20.5}
\end{eqnarray}
The corresponding star-analogues of these relations being rewritten
in components will read
\begin{eqnarray}
L^a[n+g(L^b)]\cdot L^b[n]=
R^{ab}_{cd}\ L^c[n+g(L^d)]\cdot L^d[n].
\label{20.6}
\end{eqnarray}
These relations also do not form any Lie$_q$ algebra.
We call an algebra described by them the 
quasi-Lie$_q$ algebra.

\medskip
Consider an example. In section \ref{7} we already
considered the case of quasi-$sl(2)$. Consider now its
$q$-deformed version. Let ${\cal L}=sl_q(2)$ with four generators $a,
b, c, d$ with gradings $0, -1, +1, 0$, respectively.
Then the standard defining relations for this algebra read
\begin{eqnarray}
ab&=&qba,\nonumber\\ 
cd&=&qdc,\nonumber\\ 
ac&=&qca,\nonumber\\ 
bd&=&qdb,\nonumber\\
ad-da&=&qbc-q^{-1}cb, \nonumber\\
bc&=&cb,
\label{20.7}
\end{eqnarray}
with an additional requirement
\begin{eqnarray}
\det_q \left(
\begin{array}{cc}
a&b\\
c&d
\end{array}\right)
=ad-qbc=1.
\label{20.8}
\end{eqnarray}
Using the simplified notations $a_n=a[n],
b_n=b[n], c_n=c[n], d_n=d[n]$, 
we can write down the quasi-counterparts of
relations (\ref{20.7}) 
and (\ref{20.8}) which respectively will read
\begin{eqnarray}
a_{n-1}b_n&=&qb_na_n,\nonumber\\ 
c_nd_n&=&qd_{n+1}c_n, \nonumber\\ 
a_{n+1}c_n&=&qc_na_n,\nonumber\\ 
b_nd_n&=&qd_{n-1}b_n,\nonumber\\
a_nd_n-d_na_n&=&qb_{n+1}c_n-q^{-1}c_{n-1}b_n, \nonumber\\ 
b_{n+1}c_n&=&c_{n-1}b_n
\label{20.12}
\end{eqnarray}
and
\begin{eqnarray}
a_nd_n-qb_{n+1}c_n=1.
\label{20.13}
\end{eqnarray}
We call an algebra defined 
by these relations the ``quasi-$sl_q(2)$''.

\section{The main theorem}
\label{21}

Collecting the results of sections \ref{17} -- \ref{20} we
can formulate the following theorem.

\medskip
{\bf Theorem.} Let ${\cal L}$ be a certain Lie or lie$_q$
algebra defined by commutation relations (\ref{20.3}) or (\ref{20.5}). 
Assume that its generators $L^a$ obey some polynomial relations 
\begin{eqnarray}
\sum_{j,k}\sum_{a_1,\ldots,a_k} C_{a_ka_{k-1}\ldots a_2a_1}^j 
L^{a_k}L^{a_{k-1}}\ldots L^{a_2}L^{a_1}=0
\label{21.1}
\end{eqnarray}
derivable from the defining relations (\ref{20.3}) or (\ref{20.5}) 
of algebra ${\cal L}$. 
Then the generators $L^a[n]$ 
of an abstract quasi-${\cal L}$ algebra defined 
by quasi-commutation relations (\ref{20.4}) or (\ref{20.6})
obey the ``quasi'' analogues of relations (\ref{21.1})
which read
\begin{eqnarray}
\sum_{j,k}\sum_{a_1,\ldots,a_k} C_{a_ka_{k-1}\ldots a_2a_1}^j 
L^{a_k}[n+g_{k-1}]L^{a_{k-1}}[n+g_{k-2}]\ldots 
L^{a_2}[n+g_1]L^{a_1}[n+g_0]=0
\label{21.2}
\end{eqnarray}
and in which
\begin{eqnarray}
g_0=0,\quad g_k=\sum_{i=1}^k g(L^{a_i}).
\label{21.3}
\end{eqnarray}

\medskip
{\bf Proof.} Denote by $A$ the universal enveloping algebra
of algebra ${\cal L}$. It obviously can be considered as 
a $Z^r$-graded associative algebra with unit.
Using generators $L^a[n]$ of a given quasi-${\cal L}$ algebra,
we can construct the composite operators $\hat L^a$ whose
product is defined by formula (\ref{18.1}). The 
algebra $\hat{\cal L}$ of these operators is 
isomorphic to ${\cal L}$ by construction. Let us now
construct the universal enveloping algebra $\hat A$ for
${\cal L}$. The isomorphism of algebras ${\cal L}$ and
$\hat{\cal L}$ implies the isomorphism of their
universal enveloping algebras $A$ and $\hat A$.
This, in turn, implies the existence of 
hat-versions of relations (\ref{21.1}) in $\hat A$. 
Rewriting these hat-relations in the components we obtain formulas
(\ref{21.2}) and (\ref{21.3}), which completes the proof.

\medskip
{\bf Corollary 1.} All relations in algebra ${\cal T}_R$
imply the existence of analogous quasi-relations in
the corresponding quasi-${\cal T}_R$ algebra. 
These quasi-relations can be obtained in the following simple way. 
One should consider separately each monomial in an initial
${\cal T}_R$-algebraic relation and endow all its co-factors by 
one and the same index $n\in Z^r$. 
After this one should shift the index of 
each cofactor by a total grading of all right-standing 
co-factors. Collecting all monomials transformed in
such a way together, 
we obtain the desired form of quasi-${\cal T}$-algebraic
relation.

\medskip
{\bf Corollary 2.} Each integrable and exactly solvable
model associated with algebra ${\cal T}_R$ has an
integrable and quasi-exactly solvable conterpart associated
with the corresponding quasi-${\cal T}_R$ algebra. 
Applying the rules given in corolary 1 to formulas 
describing the form of commuting integrals of motion
and determining the 
structure of Bethe ansatz solutions of integrable
and exactly solvable models, we can obtain the
corresponding formulas 
for integrable and quasi-exactly solvable models.
In particular, all formulas of sections \ref{3} -- \ref{7}
and \ref{9} -- \ref{12} can be obtained elementary from the
analogous formulas of sections \ref{2} and \ref{8}.

\medskip
From these two corollaries it immediately follows that the only 
non-triviality in constructing the 
quasi-exactly solvable modifications of
integrable exactly solvable models lies in finding the
realizations of quasi-${\cal T}_R$ algebras defined by 
quasi-commutation relations
\begin{eqnarray}
R_{\alpha\beta\gamma\delta}(\lambda-\mu)
T_{\gamma\rho}(\lambda,n+g_{\delta\sigma})T_{\delta\sigma}(\mu,n)=
T_{\beta\delta}(\mu,n+g_{\alpha\gamma})T_{\alpha\gamma}(\lambda,n)
R_{\gamma\delta\rho\sigma}(\lambda-\mu)
\label{21.4}
\end{eqnarray}
with $g_{\alpha\beta}=g(T_{\alpha\beta}(\lambda,n))$.
One of the possible ways of constructing such realizations
was described in section \ref{19}. 
It is quite obvious however that the proposed realizations
do not expire the set of all possible ones. This can be seen
even from the defining relations (\ref{3.1}) and (\ref{9.1}) 
for quasi-${\cal G}[sl(2)]$ and quasi-${\cal Y}[sl_q(2)]$
algebras which leave a considerable freedom in choosing
the form of their 
generators (see formulas (\ref{3.1a})and (\ref{9.1a})). 
The problem of listing all non-equivalent realizations
of these algebras is only a part of the following
general one:

\medskip
{\bf Problem.} Construct the representation theory for
quasi-${\cal T}_R$ algebras.

\medskip
We expect that solution of this problem may lead to
revealing new interesting mathematical structures and notions
which neccessarily find their applications in the theory of
quasi-exactly solvable systems.

\section{The classical limit of quasi-Yang-Baxter algebras}
\label{22}

It is known that the Gaudin algebra ${\cal G}[sl(2)]$ has a
well defined classical conterpart --- the so-called
classical Gaudin algebra. The corresponding 
Lie-Poisson commutation relations for its generators (which
in this case take the meaning of functions on the phase space) read
\begin{eqnarray}
\{S^0(\lambda),S^0(\mu)\}&=&0, \nonumber\\
\{S^+(\lambda),S^+(\mu)\}&=&0, \nonumber\\
\{S^-(\lambda),S^-(\mu)\}&=&0, \nonumber\\
\{S^0(\lambda),S^+(\mu)\}&=&
-(\lambda-\mu)^{-1}\{S^+(\lambda)-S^+(\mu)\},\nonumber\\
\{S^0(\lambda),S^-(\mu)\}&=&
+(\lambda-\mu)^{-1}\{S^-(\lambda)-S^-(\mu)\},\nonumber\\
\{S^-(\lambda),S^+(\mu)\}&=&
-2(\lambda-\mu)^{-1}\{S^0(\lambda)-S^0(\mu)\}
\label{22.1}
\end{eqnarray}
and imply the commutativity of functions
\begin{eqnarray}
H(\lambda)=S^0(\lambda)S^0(\lambda)-S^+(\lambda)S^-(\lambda)
\label{22.2}
\end{eqnarray}
playing the role of integrals of motion of a classical
completely integrable system. The latter is usually
reffered to as the classical Gaudin model.

\medskip
It is naturally to ask ourselves if there exists any classical
limit of the quasi-Gaudin algebra defined in section \ref{3}.
The answer to this question is positive. In order to
demonstrate this we note that the quasi-Gaudin algebra
defined by quasi-commutation relations (\ref{3.1}) admits
a class of equivalent representations which can be obtained
by an appropriate rescaling the integer parameter $n$.
Indeed, it is not difficult to see that if one uses instead
of (\ref{3.1}) the following modified relations:
\begin{eqnarray}
S_{n\hbar}^0(\lambda)S_{n\hbar}^0(\mu)-
S_{n\hbar}^0(\mu)S_{n\hbar}^0(\lambda)&=&0, \nonumber\\
S_{n\hbar+\hbar}^+(\lambda)S_{n\hbar}^+(\mu)-
S_{n\hbar+\hbar}^+(\mu)S_{n\hbar}^+(\lambda)&=&0, 
\nonumber\\
S_{n\hbar-\hbar}^-(\lambda)S_{n\hbar}^-(\mu)-
S_{n\hbar-\hbar}^-(\mu)S_{n\hbar}^-(\lambda)&=&0, 
\nonumber\\
S_{n\hbar+\hbar}^0(\lambda)S_{n\hbar}^+(\mu)-
S_{n\hbar}^+(\mu)S_{n\hbar}^0(\lambda)&=&
-\hbar(\lambda-\mu)^{-1}\{S_{n\hbar}^+(\lambda)-S_{n\hbar}^+(\mu)\},
\nonumber\\
S_{n\hbar-\hbar}^0(\lambda)S_{n\hbar}^-(\mu)-
S_{n\hbar}^-(\mu)S_{n\hbar}^0(\lambda)&=&
+\hbar(\lambda-\mu)^{-1}\{S_{n\hbar}^-(\lambda)-S_{n\hbar}^-(\mu)\},
\nonumber\\
S_{n\hbar+\hbar}^-(\lambda)S_{n\hbar}^+(\mu) -
S_{n\hbar-\hbar}^+(\mu)S_{n\hbar}^-(\lambda)&=&
-2\hbar(\lambda-\mu)^{-1}\{S_{n\hbar}^0(\lambda)-S_{n\hbar}^0(\mu)\},
\label{22.3}
\end{eqnarray}
in which $\hbar$ is a parameter, then
the operators 
\begin{eqnarray}
H_{n\hbar}(\lambda)=S_{n\hbar}^0(\lambda)S_{n\hbar}^0(\lambda)-
\frac{1}{2}
S_{n\hbar+\hbar}^-(\lambda)S_{n\hbar}^+(\lambda)-
\frac{1}{2}S_{n\hbar-\hbar}^+(\lambda)S_{n\hbar}^-(\lambda)
\label{22.4}
\end{eqnarray}
will as before commute with each other for any given $n$.
Let us now rewrite formula (\ref{22.3}) in the form
\begin{eqnarray}
\frac{[S_{n\hbar}^0(\lambda),S_{n\hbar}^0(\mu)]}{\hbar}&=&0, \nonumber\\
\frac{[S_{n\hbar}^+(\lambda),S_{n\hbar}^+(\mu)]}{\hbar}+
\frac{S_{n\hbar+\hbar}^+(\lambda)-S_{n\hbar}^+(\lambda)}
{\hbar}S_{n\hbar}^+(\mu)-
\frac{S_{n\hbar+\hbar}^+(\mu)-S_{n\hbar}^+(\mu)}
{\hbar}S_{n\hbar}^+(\lambda)&=&0, \nonumber\\
\frac{[S_{n\hbar}^-(\lambda),S_{n\hbar}^-(\mu)]}{\hbar}+
\frac{S_{n\hbar-\hbar}^-(\lambda)-S_{n\hbar}^-(\lambda)}
{\hbar}S_{n\hbar}^-(\mu)-
\frac{S_{n\hbar-\hbar}^-(\mu)-S_{n\hbar}^-(\mu)}
{\hbar}S_{n\hbar}^-(\lambda)&=&0, \nonumber\\
\frac{[S_{n\hbar}^0(\lambda),S_{n\hbar}^+(\mu)]}{\hbar}+
\frac{S_{n\hbar+\hbar}^0(\lambda)-S_{n\hbar}^0(\lambda)}
{\hbar}S_{n\hbar}^+(\mu)+
\frac{S_{n\hbar}^+(\lambda)-S_{n\hbar}^+(\mu)}{\lambda-\mu}&=&0,
\nonumber\\
\frac{[S_{n\hbar}^0(\lambda),S_{n\hbar}^-(\mu)]}{\hbar}+
\frac{S_{n\hbar-\hbar}^0(\lambda)-S_{n\hbar}^0(\lambda)}
{\hbar}S_{n\hbar}^-(\mu)-
\frac{S_{n\hbar}^-(\lambda)-S_{n\hbar}^-(\mu)}{\lambda-\mu}&=&0,
\nonumber\\
\frac{[S_{n\hbar}^-(\lambda),S_{n\hbar}^+(\mu)]}{\hbar} +
\frac{S_{n\hbar+\hbar}^-(\lambda)-S_{n\hbar}^-(\lambda)}
{\hbar}S_{n\hbar}^+(\mu)-
\frac{S_{n\hbar-\hbar}^+(\mu)-S_{n\hbar}^+(\mu)}
{\hbar}S_{n\hbar}^-(\lambda)&=&\nonumber\\
=-2\frac{S_{n\hbar}^0(\lambda)-S_{n\hbar}^0(\mu)}{\lambda-\mu}
\label{22.5}
\end{eqnarray}
and apply to it a standard dequantization procedure. For
this one should assume that the parameter $\hbar$ is small,
$n$ is large, and their product $t=n\hbar$ is finite. Considering $t$ as
a new continuous variable and taking the limits $\hbar\rightarrow 0$ and
$n\rightarrow\infty$, we can replace the 
commutatators in (\ref{22.5}) by Poisson brackets and the finite
differences by derivatives. The result will read
\begin{eqnarray}
\{S_{t}^0(\lambda),S_{t}^0(\mu)\}&=&0, \nonumber\\
\{S_{t}^+(\lambda),S_{t}^+(\mu)\}+
\dot S_{t}^+(\lambda)S_{t}^+(\mu)-
\dot S_{t}^+(\mu)S_{t}^+(\lambda)&=&0, \nonumber\\
\{S_{t}^-(\lambda),S_{t}^-(\mu)\}-
\dot S_{t}^-(\lambda)S_{t}^-(\mu)-
\dot S_{t}^-(\mu)S_{t}^-(\lambda)&=&0, \nonumber\\
\{S_{t}^0(\lambda),S_{t}^+(\mu)\}+
\dot S_{t}^0(\lambda)S_{t}^+(\mu)+
\frac{S_{t}^+(\lambda)-S_{t}^+(\mu)}{\lambda-\mu}&=&0,
\nonumber\\
\{S_{t}^0(\lambda),S_{t}^-(\mu)\}+
\dot S_{t}^0(\lambda)S_{t}^-(\mu)-
\frac{S_{t}^-(\lambda)-S_{t}^-(\mu)}{\lambda-\mu}&=&0,
\nonumber\\
\{S_{t}^-(\lambda),S_{t}^+(\mu)\} +
\dot S_{t}^-(\lambda)S_{t}^+(\mu)-
\dot S_{t}^+(\mu)S_{t}^-(\lambda)+
2\frac{S_{t}^0(\lambda)-S_{t}^0(\mu)}{\lambda-\mu}&=&0,
\label{22.6}
\end{eqnarray}
where the dot denotes the derivative with respect to $t$.
We call an algebra defined by these relations 
the classical quasi-Gaudin algebra. The existence of this
algebra can easily be proved if one replaces $n$ by
$n\hbar$ in formulas (\ref{5.2}), takes the limits
$\hbar\rightarrow 0$ and $n\rightarrow\infty$, and checks
that the resulting generators (having the same form
as in (\ref{5.2}) but with $n$ replaced by $t$) obey the
relations (\ref{22.6}). Taking the same limits 
in formula (\ref{22.4}), we obtain
\begin{eqnarray}
H_{t}(\lambda)=S_{t}^0(\lambda)S_{t}^0(\lambda)
-S_{t}^+(\lambda)S_{t}^-(\lambda).
\label{22.7}
\end{eqnarray}
It is not difficult to see that functions (\ref{22.7})
are in involution for any given $t$,
\begin{eqnarray}
\{H_{t}(\lambda), H_{t}(\mu)\}=0,
\label{22.8}
\end{eqnarray}
but this is generally not true if the values of
$t$ differ from each other. 
This means that, exactly as in the quantum case, the
classical quasi-Gaudin algebra generates an infinite number
(or, more exactly, a continuous family) of integrable
classical systems. It is natural to call them the classical
quasi-Gaudin models. 

\medskip
According to the famous Liouville theorem, all
these models are integrable in quadratures and, in this
sense, can be qualified as exactly solvable. Remember,
however, that the practical usefulness of the Liouville
theorem is very low if one deals with classical systems with
many degrees of freedom. In general, we simply cannot present their
explicit solutions!\footnote{This situation is exactly the
same as in quantum mechanics: we become happy when a
certain quantum problem is reduced to a purely algebraic one and
call such a problem exactly solvable (or quasi-exactly solvable).
But it does not mean that we are able to solve 
any algebraic problem explicitly, especially if the number
of unknowns is large.} At present time there is only one
constructive way for solving classically integrable systems.
It is given by the famous classical inverse scattering
method (CISM) (see e.g. \cite{FaTa87})
which presents more or less explicit algorithm of
constructing action-angle variables for models associated
with classical Yang-Baxter algebras\footnote{The fact that
CISM works for integrable models associated with classical
Yang-Baxter algebras does not mean that it works for any
integrable model.}. We expect that the classical 
quasi-Gaudin models which we
obtained are in fact quasi-exactly solvable in the sense of
CISM. This means that using
the standard prescriptions of this method and applying them
to models (\ref{22.7}) we probably will obtain only some
part of action-angle variables, but not a complete set of
them. It would be very desirable to try to verify this conjecture 
because if it is true then one obtains obtain a remarkable 
possibility of introducing a concept
of quasi-exact solvability in classical mechanics.

\medskip
Note that the relations (\ref{22.6}) can be rewritten in
more standard form if one introduces a new bracket:
$\{\{\ ,\}\}$. Defining it as
\begin{eqnarray}
\{\{a_t,b_t\}\}=\{a_t,b_t\}+g(a_t)\dot a_tb_t
-g(b_t)\dot b_ta_t,
\label{22.9}
\end{eqnarray}
where $g(a)$, as usually, denotes a grading, 
we obtain for (\ref{22.6}): 
\begin{eqnarray}
\{\{S^0(\lambda),S^0(\mu)\}\}&=&0, \nonumber\\
\{\{S^+(\lambda),S^+(\mu)\}\}&=&0, \nonumber\\
\{\{S^-(\lambda),S^-(\mu)\}\}&=&0, \nonumber\\
\{\{S^0(\lambda),S^+(\mu)\}\}&=&
-(\lambda-\mu)^{-1}\{S^+(\lambda)-S^+(\mu)\},\nonumber\\
\{\{S^0(\lambda),S^-(\mu)\}\}&=&
+(\lambda-\mu)^{-1}\{S^-(\lambda)-S^-(\mu)\},\nonumber\\
\{\{S^-(\lambda),S^+(\mu)\}\}&=&
-2(\lambda-\mu)^{-1}\{S^0(\lambda)-S^0(\mu)\}.
\label{22.10}
\end{eqnarray}
We see that the classical quasi-Gaudin
algebra can be considered as an ordinary classical Gaudin
algebra realized with respect to the non-standard bracket (\ref{22.9}).
We arrived at the same situation which we had in section
\ref{18} for quasi-commutators. This similarity is not accidental
because the quasi-Poisson bracket (\ref{22.9}) is nothing
else than the classical 
limit of a quasi-commutator which can be defined as
\begin{eqnarray}
[[a_n,b_n]]=a_{n+g(b)}b_n-b_{n+g(a)}a_n.
\label{22.11}
\end{eqnarray}
Indeed, following the standard prescriptions of 
dequantization procedure, 
we should rescale the relation (\ref{22.11}) as
\begin{eqnarray}
[[a_{n\hbar},b_{n\hbar}]]=a_{n\hbar+g(b)\hbar}b_{n\hbar}-
b_{n\hbar+g(a)\hbar}a_{n\hbar}
\label{22.12}
\end{eqnarray}
and then rewrite it in the form
\begin{eqnarray}
[[a_{n\hbar},b_{n\hbar}]]=[a_{n\hbar},b_{n\hbar}]+
\hbar\frac{a_{n\hbar+g(b)\hbar}-a_{n\hbar}}{\hbar}b_{n\hbar}-
\hbar\frac{b_{n\hbar+g(a)\hbar}-b_{n\hbar}}{\hbar}a_{n\hbar}
\label{22.13}
\end{eqnarray}
Taking now the limits $\hbar\rightarrow 0$ and
$n\rightarrow\infty$, introducing a new finite and
continuous variable 
$t=n\hbar$ and using definition (\ref{22.9}), we obtain
\begin{eqnarray}
[[a_{n\hbar},b_{n\hbar}]]\rightarrow \hbar
\{\{a_{t},b_{t}\}\}.
\label{22.14}
\end{eqnarray}
From this analysis it follows that the definition 
(\ref{22.9}) works not only for the
objects with gradings $0$ and $\pm 1$ but for arbitrarily
graded ones. One can check that the quasi-Poisson bracket is
anti-symmetric 
\begin{eqnarray}
\{\{a_t,b_t\}\}+\{\{b_t,a_t\}\}=0
\label{22.15}
\end{eqnarray}
and obeys the standard Jacoby identity:
\begin{eqnarray}
\{\{\ \{\{a_t,b_t\}\},c_t\}\}+\{\{\ \{\{b_t,c_t\}\},a_t\}\}
+\{\{\ \{\{c_t,a_t\}\},b_t\}\}=0.
\label{22.16}
\end{eqnarray}
For zero-graded objects it coincides with an ordinary
Poisson bracket
\begin{eqnarray}
\{\{a_t,b_t\}\}=\{b_t,a_t\},\quad \mbox{if} \ g(a)=g(b)=0
\label{22.17}
\end{eqnarray}
and therefore the commutativity of functions $H_t(\lambda)$
can be considered as an elementary consequene of the 
commutativity of their Gaudin analogues $H(\lambda)$.

\section{The perspectives}
\label{23}

We completed the exposition of the proposed $R$-matrix approach
to the problem of quasi-exact solvability. Summarizing, we
can say that we found a 
simple modification of the celebrated quantum inverse 
scattering method which leads to new wide classes of completely 
integrable models. The latter are distinguished by the fact they 
admit an algebraic Bethe ansatz solution only for some limited
parts of the spectrum, but not for the whole spectrum, as in
the standard version of QISM. Therefore they are typical
quasi-exactly solvable models discussed in refs. \cite{Us88b,Us89,Us92}. 
An underlying algebra responsible for both the phenomena of complete 
integrability and quasi-exact solvability was constructed.
We called it the ``quasi-${\cal T}_R$ algebra'' and demonstrated
that it can be considered as a very simple (but non-trivial) 
deformation of the ordinary ${\cal T}_R$ algebra of monodromy 
matrices generated by various solutions of Yang--Baxter equation. 
This enabled one to claim that we actually have found a new
class of fundamental symmetries of integrable quantum systems.
The main attention in the paper was devoted 
to construction of such deformations for simplest ${\cal
T}_R$ algebras, i.e. for Gaudin algebra ${\cal G}[sl(2)]$
and its quantized version --- the Yangian ${\cal Y}[sl(2)]$.
We also constructed and solved the quasi-exactly solvable models of 
magnetic chains generated by these algebras in the
framework of QISM. The classical vesions of quasi-${\cal
T}_R$ algebras were discussed briefly and we intend to
return to this interesting question in the nearest future.

\medskip
The main result of this paper lies however in the proof of the
fact that any graded associative algebra with unit admits
such ``quasi'' deformation. We showed that this deformation modifies
the structure of the elements of an algebra and changes the
definition of their associative product. At the same time,
despite the numerous attempts we did not manage to find any
appropriate deformation of a co-product which would
preserve a possible Hopf algebraic structure of the initial
algebra. This enables one to make a conjecture that
quasi-${\cal T}_R$ algebras which we constructed in this
paper {\it are not} the Hopf algebras\footnote{It would be
nice to call such algebras ``quasi-Hopf algebras'', but, unfortunately,
this term is already used in a different sense \cite{Dr89}.}. 

\medskip
The method we proposed in this paper is interesting
primarily because it opens a practical way of constructing 
quasi-exactly solvable deformations of {\it all} known exactly
solvable and completely integrable quantum models
obtainable in the framework of standard version of QISM.
The list of such models includes: 

\medskip
1) Exactly solvable models of ordinary quantum mechanics.
In one-dimensional case these are: 
the simple harmonic oscillator, the Kratzer and Morse potentials, 
P\"oschel--Teller potential wells, etc. In multi-dimensional case 
these are: the multi-dimensional harmonic oscillator and also
various separable models defined on non-trivial curved
manifolds and constructed in ref. \cite{Us94}. As was demonstrated
in \cite{Us94}, all these models can be considered as
various degenerate cases of Gaudin models and hence can be
obtained in the framework of classical $R$-matrix approach.
This, in turn, means that we can apply to these models our
deformation procedure and obtain their quasi-exactly
solvable conterparts. This would be a perfect test for the
proposed method because the explicit form of all these
quasi-exactly solvable models is already known (see
e.g. refs. \cite{Us88a,Tu88,Us89,Us94}).

\medskip
2) Exactly solvable models of multi-particle
quantum mechanics. These are: the Calogero model, the Calogero-Moser
model and other 
similar models discussed for example in ref. \cite{OlPe83}. 
Recently it was demonstrated that these models
can be obtained in the framework of the so-called method of dynamical 
$R$-matrices. A main distinguished feature of this method lies in the
fact that the $R$-matrices used in it depend themselves on 
dynamical variables, i.e. on the components of a monodromy matrix.
For us however it does not matter what kind of
$R$-matrices is used in commutation relations of algebra
${\cal T}_R$ because of the universality of the proposed deformation
method. This means that we can use this method for constructing
quasi-exactly solvable versions of multi-particle Calogero
and Calogero-Moser models. Note that particular classes of
such models have already been constructed in ref.
\cite{Us91} in different way (see also the book \cite{Us94}).

\medskip
3) Exactly solvable models of spin chains. These include,
first of all, the local chains: the simples $1/2$-spin
chains with $XXX$, $XXY$ and $XYZ$ symmetries as well as their
higher spin generalizations constructed in refs. \cite{TaTaFa83,Ta85},
and also the various inhomogeneous spin chains with the same symmetries.
These models are associated respectively with rational,
trigonometric and elliptic solutions of Yang--Baxter equation.
In this paper we discussed only a very particular case of
$XXX$ models. Moreover, for these models we presented only
the sets of 
commuting integrals of motion but did not discuss the methods of
extracting from these sets the hamiltonians of physically
interesting systems. 

\medskip
4) Exactly solvable models of field theory. We know a lot
of models of such a sort including  non-relativistic models
like non-linear Schr\"odinger equation, sine-Gordon model,
various lattice models, and also the relativistic models
describing various forms of four-fermion interaction.
Using our deformation technique, one can try to construct
their quasi-exactly solvable conterparts. This, may be, requires
some comments. The quasi-exactly solvable problems which were
considered up to now were defined as problems having a
certain finite parts of the spectrum constructable in an
algebraic way. If the number of such algebraically
constructable states is $N$, then, in order to find them, 
one should solve a certain $N$-th order algebraic equation.
A fruitfulness of such a concept in quantum mechanics is very well known.
However, if we want to have something like field-theoretical 
quasi-exactly solvable models (or, simply, quasi-exactly solvable
models with many degrees of freedom)
one should allow $N$ to tend to infinity, because the only
physically interesting 
situations in such models may appear when the number
of excitations is infinitely (or, simply, very) large.
But for this we should be able to solve the corresponding 
``algebraic'' equations! If we speak only of the
algebraizability itself, as, for example, in the partial 
algebraization method, then we are unable
to solve this problem, since the complexity of
corresponding algebraic problems catastrophically increases
with the increase of $N$. However, if we have some
additional structure in our systems, which is, say,
the integrability property, then, as we have seen above, the
equations for solving our algebraic problem take the form
of Bethe ansatz equations which in the large $N$ limit are
well defined and have been studied in detail in many occasions.

\medskip
5) Exactly solvable models of statistical mechanics. We
know that the mathematical structure lying behind such
models is again the ${\cal T}_R$ algebra, which however has
now the meaning of an algebra of transfer matrices. It
would be very interesting to try to understand what kind of
statistical systems correspond to quasi-${\cal T}_R$ algebras,
and in which sense they are quasi-exactly solvable.

\medskip
Of course, the construction of quasi-exactly solvable models with
an infinite number of degrees of freedom in the framevork of 
quasi-QISM approach is a long-term program which
hardly can be solved in the nearest future. Now we can only
express an optimistic hope that its realization is possible. The
reason for our optimism is based on two facts: a) that the
standard QISM formalism is well adopted to the 
study of exactly solvable models with many
degrees of freedom, and b) that there is no essential difference
between QISM and quasi-QISM, as it follows from the results
of our paper. Of course, the quasi-QISM does not exactly
coincide with the standard QISM and has many specific features 
which need a very careful study and understanding. In this connection,
I would like to mention two important problems which
immediately arise if one wants to give a physical
interpretation of the results of the present paper.

\medskip
{\bf The hermiticity problem.} Any physically realistic
model should be described by hermitian hamiltonians. If one
looks at quasi-Gaudin and quasi-$XXX$ models (\ref{6.1}) and 
(\ref{12.1}) which we constructed in our paper, then it
becomes obvious that their hermiticity properties strongly
depend on the choice of a representation in which the
corresponding ${\cal G}[sl(2)]$ and ${\cal Y}[sl(2)]$
algebras act. One of trivial possibilities of making these hamiltonians
hermitian and the corresponding wavefunctions normalizable
is to take the unitary representations of these algebras which,
due to their incompactness, are all infinite-dimensional.
Are there any other ways? The experience accumulated in the
study of integrable and quasi-exactly solvable models
enables one to answer this question positively. Indeed, there is 
a constructive method for building such models 
(and this is just the inverse method of separation of variables
discussed in section \ref{15}) 
which allows one to guarantee the hermiticity of the
resulting hamiltonians from very beginning, i.e. by construction
(for more detail see e.g. ref. \cite{Us92,Us94}). 
Unfortunately, in order to apply this method to models (\ref{6.1})
and (\ref{12.1}) as well as to other models obtainable in
the framework of quasi-QISM, one should first separate the
variables in them (we mean here the separation of variables 
in the generalized Sklyanin sense \cite{Sk95}). This immediately creates
two important questions: a) Is there any regular way for
performing Sklyanin's procedure of separation of variables
for models obtainable in the framework of quasi-QISM? 
b) Is there any direct way of formulating a criterion of
hermiticity immediately in terms of quasi-${\cal T}_R$ algebras and
their representations? The second question creates, in turn,
a problem of developing some kind of representation theory 
for quasi-${\cal T}_R$ algebras, etc. 

\medskip
{\bf The problem of locality.} This is no less important
problem on whose positive solution will finally depend a
progress in the realization of our program. We know that most of
physically interesting models of field theory are local.
At the same time the models (\ref{6.1}) and (\ref{12.1})
look as typically non-local ones. Then the following question
immediately arises: Is it possible to extract from the set
of integrals of motion of these models some local hamiltonians?
And, if not, then may be there exist other realizations of 
quasi-${\cal G}[sl(2)]$ and quasi-${\cal Y}[sl(2)]$
algebras for which it would be possible?

\medskip
I hope that the hermiticity and locality problems will
finally find their solutions in the framework of quasi-QISM approach.
The more that the construction of hermitian and
local quasi-exactly solvable models with many degrees of
freedom is possible even in the standard version of QISM.
I am addressing the interested reader to ref. \cite{Us96} in which this
question is discussed in detail.

\section*{Acknowledgements}

I would like to thank dr. T. Brzezinski and profs. 
S. Novikov, J. Wess and P. Wiegmann
for interesting discussions. This work was
partially supported by DFG grant No. 436 POL 113/77/0 (S).

\end{document}